\documentclass[12pt]{article}

\usepackage[utf8]{inputenc} 
\usepackage{amsmath, amssymb} 
\usepackage{graphicx} 
\usepackage{hyperref} 
\usepackage{geometry} 
\usepackage{float}
\usepackage{caption}
\usepackage{subcaption}
\usepackage{physics}
\usepackage{comment}
\usepackage{xcolor}
\usepackage{authblk}

\title{Finite-Time Protocols Stabilize Charging in Noisy Ising Quantum Batteries}

\author[1,2]{Riccardo Grazi}
\author[3]{Henrik Johannesson}
\author[1,2]{Dario Ferraro}
\author[1,2]{Niccolò Traverso Ziani}

\affil[1]{\small Dipartimento di Fisica, Università degli Studi di Genova, via Dodecaneso 33, 16146, Genova, Italy}
\affil[2]{\small CNR SPIN, via Dodecaneso 33, 16146, Genova, Italy}
\affil[3]{\small Department of Physics, University of Gothenburg, Gothenburg, SE 412 96, Sweden.}
\date{}
\begin{document}

\maketitle

\begin{abstract}
Reliable charging protocols are crucial for advancing quantum batteries toward practical use. We investigate a transverse-field Ising chain as a quantum battery, focusing on the combined role of qubit interactions in the battery model and finite charging time. This interplay yields smoother and more controllable charging compared to sudden protocols or non-interacting batteries. Introducing stochastic noise reveals a strong dependence on the charging trajectory. Protocols that weakly excite the system gain energy under noise but lose extractable work. In contrast, protocols that strongly excite many modes show the opposite trend: noise reduces stored energy yet improves efficiency, defined as the ratio of ergotropy to stored energy. These findings demonstrate that finite-time ramps stabilize charging and highlight that noise can either hinder or enhance quantum-battery performance depending on the protocol.
\end{abstract}

\section{Introduction}
Quantum batteries (QBs)~\cite{Alicki13, Quach23, Campaioli24} have emerged as a promising platform at the intersection of quantum thermodynamics~\cite{Vinjanampathy16, Bhattacharjee21, Deffner19}, quantum materials~\cite{Keimer17, Giustino20}, and quantum technologies~\cite{Aguado24}. They provide a testbed for exploring how coherence, correlations, and collective dynamics ~\cite{Binder15, Campaioli17, Gyhm22} can be harnessed to enhance charging power~\cite{Ferraro18, Rossini20}, storage capacity~\cite{Ferraro19, Julia20}, energy extraction efficiency~\cite{Barra19, Hovhannisyan20, Caravelli21, Cheng25, Razzoli25, Cavaliere25, Beder25} and operational precision~\cite{Friis18, Rinaldi24, Donelli25}. Several figures of merit have been introduced to characterize QB performance, including stored energy~\cite{Andolina18}, charging power, and ergotropy~\cite{Allahverdyan04}—the maximum extractable work under unitary operations. Within this broad landscape, several physical systems have been analyzed as quantum batteries, including the solid-state Dicke model~\cite{Ferraro18, Gemme23, Erdman24}, molecules in an organic microcavity~\cite{Quach22, Hymas25}, quantum dots~\cite{Wenniger23}, nuclear spins~\cite{Joshi22PRA, Cruz22} and spin chains~\cite{Le18, Barra22, Catalano24, Zhang24, Chand25}. Spin-chain models stand out as a minimal yet nontrivial framework, particularly since they are experimentally realizable in platforms such as Rydberg-atom arrays~\cite{Adams19, Browaeys20, Wu_2021}, superconducting qubits~\cite{Zippilli15, Geller18}, and trapped-ion crystals~\cite{Porras2004, Senko15, hess2017}.

Charging a QB requires transferring energy into the system. Such transfer can be achieved in several ways, including coupling the battery to a cavity~\cite{Ferraro18}, by Floquet driving~\cite{Zhang19}, by monitoring the systems~\cite{Elyasi25}, or through a quantum quench~\cite{Mitra18} in which a Hamiltonian parameter is varied in time. Sudden quenches have been analyzed in Jordan–Wigner solvable systems~\cite{Rossini20b, Grazi24, Grazi25, Grazi25Energies}, revealing a characteristic evolution: oscillations with sharp maxima at early times, a plateau from interference among excitation frequencies, and eventual recurrences due to finite-size effects. To achieve stability in stored energy, charging must be stopped within the plateau region, which reduces charging power. This limitation motivates the exploration of finite-time ramps~\cite{Porta20}, where the parameter change occurs over a controllable interval, as a mean to stabilize charging dynamics at short times. Moreover, realistic devices are inevitably subject to noise, raising the question of whether such protocols remain robust under non-unitary dynamics~\cite{Jafari24, Jafari25}.

In this work we investigate a transverse-field Ising chain as a QB with two main goals. First, we compare sudden quenches with finite-time ramps to assess whether smoother charging dynamics can stabilize the stored energy at short times, finding that ramps indeed suppress oscillations and allow reliable exploitation of early-time dynamics. Second, we study the impact of stochastic noise superposed on the ramp. We find that battery performance depends strongly on the quench trajectory: protocols that weakly populate Jordan–Wigner fermion states gain energy under noise but lose extractable work, whereas protocols that strongly excite many modes show the opposite trend—noise reduces stored energy yet improves efficiency, defined as the ratio of ergotropy to stored energy.

\section{Results}
\subsection{Model}
As a model, we consider a quantum Ising chain with periodic boundary conditions and a time-dependent transverse field $h(t)$. The Hamiltonian reads as
\begin{equation}
    H_{0}(t) = -\sum_{j=1}^{N} \sigma_j^x\,\sigma_{j+1}^x - h(t)\sum_{j=1}^N \sigma_j^z \label{ISING}
\end{equation}
where an overall energy scale has been omitted for convenience, $\sigma^{x/z}_{j}$ are the Pauli matrices of the $j$-th spin in the usual representation and $N$ denoting the total number of spins. The external field is given by
\begin{equation}
h(t) =
\begin{cases}
h_i + v\,t ,    & 0 \le t \le t_f,\\
h_f                      & t >t_{f}
\end{cases}
\label{ramp-quench}
\end{equation}
and it will define the charging protocol of the QB. Here, $h_i$ and $h_f$ are the pre- and post- quench fields and $v$ parametrizes the slope of the ramp. At every time, the Hamiltonian can be mapped onto a set of independent fermionic two-level systems, one for each quasimomentum mode \(k = \pm (2l-1)\pi/N\) ($l = 1,2,...,N/2$), by first applying a Jordan–Wigner transformation to the spin operators and then moving to Fourier space~\cite{Franchini16} (note that without loss of generality we have restricted to the even fermion number case). The quantum Ising Hamiltonian hence becomes
\begin{equation}
    H_0(t) = \sum_{k > 0} (c^\dagger_k ~~ c_{-k})~ H_{0,k}(t)~ (c_k ~~ c^\dagger_{-k})^T \label{IsingHam}
\end{equation}
where
\begin{equation}
    H_{0,k}(t) =  \begin{pmatrix}
        2\left(h(t)-\cos(k)\right) & 2\sin(k) \\
        2\sin(k) & -2\left(h(t)-\cos(k)\right)
    \end{pmatrix},
    \label{Hamiltonian_Per_Mode}
\end{equation}
with $c_k$ and $c^\dagger_k$ annihilation and creation operators for a fermion labelled by quasimomentum $k$. After diagonalizing $H_{0,k}(t)$, we obtain the istantaneous spectrum of the model
\begin{equation}
    \varepsilon^{\pm}_{k}(t) = \pm2\sqrt{\bigl[h(t) - \cos (k)\bigr]^{2} \;+\; \sin^2 k}\,,
\end{equation}
and the corresponding instantaneous eigenstates
\begin{eqnarray} \label{eigenstates}
\begin{cases} \ket{\chi^-_k(t)} = a_k(t) \ket{1} + b_k(t) \ket{0} \\
 \ket{\chi^+_k(t)} = -b_k(t)\ket{1} + a_k(t) \ket{0}, \\
\end{cases}
\end{eqnarray}
where $\ket{1} = (1~~0)^T, \, \ket{0} = (0~~1)^T$, and $a_k(t)= \left(2(h(t)\! -\! \cos(k)) \!-\! \varepsilon_k(t)\right)/N_k(t),\,  b_k(t)=2\sin(k)/N_k(t)$ with $N_k(t)$ normalization constants.
We recall that for time-independent fields, $h(t)=h_0$, the quantum Ising chain exhibits quantum phase transitions in the thermodynamic limit when $h_0=\pm 1$, with paramagnetic phases for $|h_0| > 1$ and a ferromagnetic phase for $|h_0| < 1$ \cite{Pfeuty1970}.

We now define
\begin{equation}
    H_{Batt} \equiv H_0(0)
\end{equation}
since we want to study a QB described by the Hamiltonian of Eq. \eqref{IsingHam} with fixed transverse field $h(t) = h(0) = h_i$. Furthermore, we take the ground states of the mode Hamiltonians in Eq. \eqref{Hamiltonian_Per_Mode} as initial states, defining $\rho_{0,k}(0) = \ket{\chi^-_k(0)}\bra{\chi^-_k(0)}$. The dynamics of each mode is described by the von Neumann equation
\begin{equation}
    \dot{\rho}_{0,k}(t) = -i[H_{0,k}(t), \rho_{0,k}(t)]. \label{VonNeumann}
\end{equation}
\subsection{Noiseless charging of a single qubit}
\begin{figure}[H]
    \centering
    \includegraphics[width=0.8\linewidth]{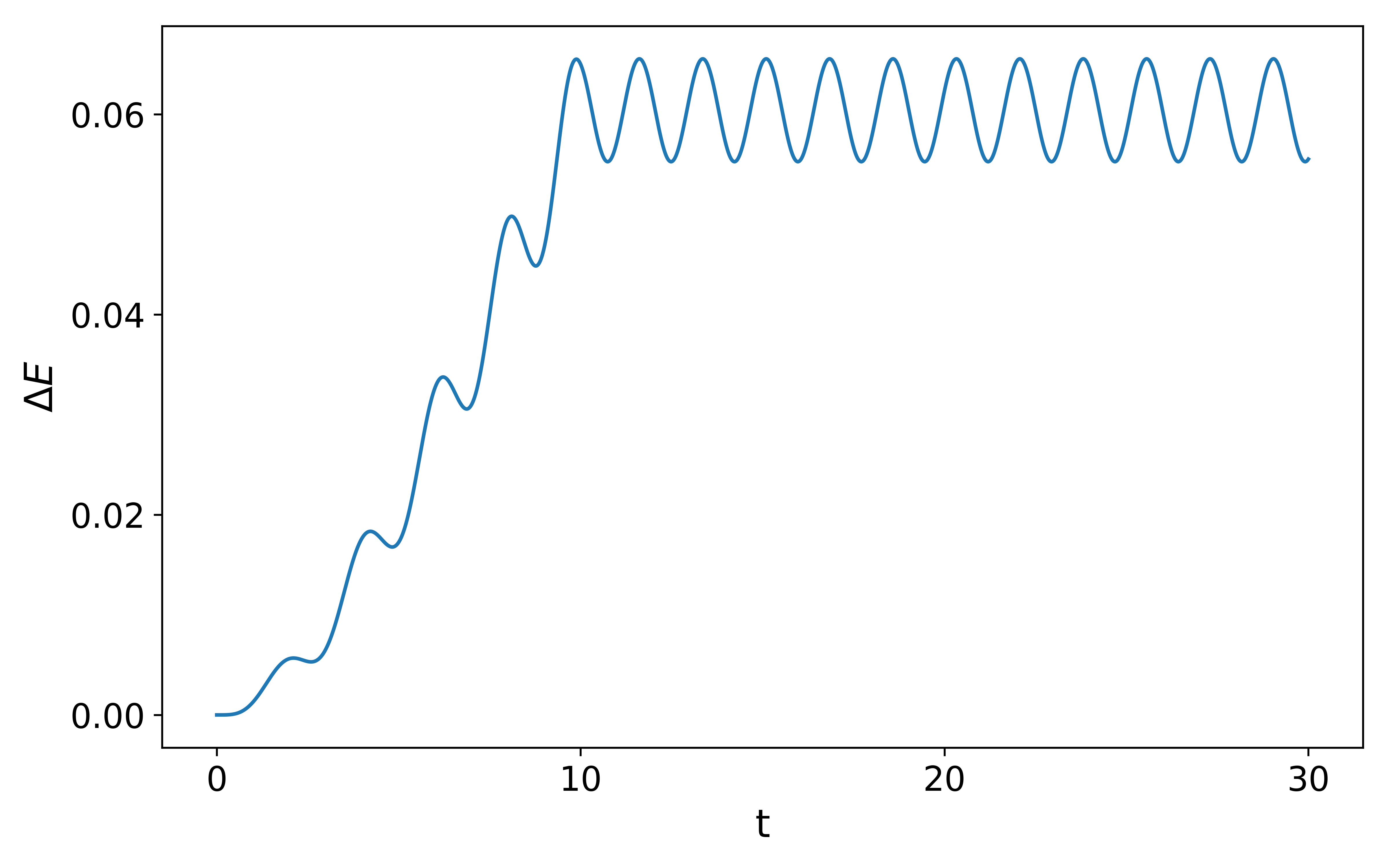}
    \caption{Energy stored in a single qubit with a charging ramp going from $t = 0$ to $t_f = 10$  as a function of time. The initial and final values of the external field are $h_i=0.8$ and $h_f=1.5$, respectively (quench rate $v=0.07$).}
    \label{DeltaESingoloQubitPlot}
\end{figure}

Since at each quasimomentum $k$ the quantum Ising chain behaves as a two-level system, it is instructive to first analyze the single-qubit Hamiltonian
\begin{equation}
    H(t) = -\sigma^x - h(t) \sigma^z
    \label{H_t}
\end{equation}
with $h(t)$ as in Eq. \eqref{ramp-quench}. For transparency and ease of notation, note that in this section we replace the mode Hamiltonian $H_{0,k}(t)$ in Eq. \eqref{Hamiltonian_Per_Mode} by the simpler single-qubit Hamiltonian as written in Eq. \eqref{H_t}. Also note that, in Eq. \eqref{H_t}, 
\begin{equation}
    H_{B} \equiv H(t=0) = - \sigma^x - h_i \sigma^z= \begin{pmatrix}
        -h_i & -1 \\
        -1 & h_i
    \end{pmatrix}
\end{equation}
represents the Hamiltonian of the single-qubit QB, while the remaining part is associated to the time dependent action of an external classical charger~\cite{Zhang19, Crescente20b}. The normalized eigenstates of $H_B$ are
\begin{equation}
    \ket{\pm} \equiv \frac{1}{\sqrt{2\omega_B(\omega_B\mp h_i)}} \begin{pmatrix}
     h_i \mp  \omega_B   \\   1
    \end{pmatrix} \label{Eigenstates_HB}
\end{equation}
having energy $\pm \omega_B$ respectively, with $\omega_B \equiv \sqrt{1+h_i^2}$. If we write the $(i,j)$-th element of the time-dependent density matrix $\rho(t)$ describing the state of the system in the basis of the eigenstates of $\sigma^z$ as $\rho_{ij}(t)$, then we can compute
\begin{equation}
    E(t)=\text{Tr}[\rho(t)H_{B}] = -h_i\left(\rho_{00}(t) - \rho_{11}(t)\right) - \left(\rho_{01}(t) +\rho_{10}(t)\right),
    \label{uxy}
\end{equation}
which leads to the stored energy~\cite{Andolina18}
\begin{equation}
    \Delta E(t) = \text{Tr}[\left(\rho(t)-\rho(0)\right)H_{B}]=E(t)-E(0).
    \label{stored}
\end{equation}
To proceed further, we consider as initial state of the QB the normalized ground state $\ket{-}$ reported in Eq. \eqref{Eigenstates_HB}. From this state we obtain the initial density matrix
\begin{equation}
    \rho(0) = \ket{-}\bra{-} = \frac{1}{2\omega_B(\omega_B+h_i)} \begin{pmatrix}
       (\omega_B + h_i)^2  & \left(\omega_B + h_i\right)\\ \left(\omega_B + h_i\right) & 1
       \label{rho_zero}
    \end{pmatrix}.
\end{equation}
Its time-evolution is simply given by the von Neumann equation
\begin{equation}
    \dot{\rho}(t) = -i[H(t), \rho(t)]
\end{equation}
which leads to the following differential equations for the elements of the qubit's density matrix
\begin{equation}
    \begin{cases}
        &\dot{\rho}_{00} = -i(\rho_{01} - \rho_{10}) \\
        &\dot{\rho}_{01} = -i\left[\left(\rho_{00} - \rho_{11}\right) - 2h(t)\rho_{01}\right]\\
        &\dot{\rho}_{10} = -i\left[\left(\rho_{11} - \rho_{00}\right) + 2h(t)\rho_{10}\right] \\
        &\dot{\rho}_{11} = -i(\rho_{10} - \rho_{01}),
    \end{cases} \label{ODEs_singoloqubit}
\end{equation}
where, to simplify the notation, we have dropped the time argument of the density matrix elements. To solve these equations, it is convenient to introduce the functions
\begin{equation}
    \begin{cases}
        &u(t) \equiv \rho_{00} - \rho_{11} \\
        &x(t) \equiv \rho_{01} + \rho_{10}\\
        &y(t) \equiv -i(\rho_{01} - \rho_{10}),
    \end{cases} \label{FuncsUXY}
\end{equation}
in such a way that the system in Eq. \eqref{ODEs_singoloqubit} becomes
\begin{equation}
    \begin{cases}
        &\dot{u} = 2y\\
        &\dot{x} = -2h(t)y\\
        &\dot{y} = -2u + 2h(t)x
    \end{cases} \label{New_Eqs}
\end{equation}
and the expression for the stored energy can be written as
\begin{equation}
     \Delta E(t) = -h_{i}\left[u(t)-u(0) \right]-\left[x(t)-x(0) \right]. \label{Energy_u_x}
\end{equation}
The detailed analysis of the single two-level system is provided in the Supporting Material. The main features, as reported in Fig.~\ref{DeltaESingoloQubitPlot}, are that the charging happens on the time scale set by the ramp, with strong oscillations with the time scale of the inverse of the relevant energies. Moreover, at fixed $h_i$ and $h_f$, the amplitude of the oscillations decreases as the ramp duration increases (more details can be found in the Supporting Material).
\subsection{Noiseless charging of the quantum Ising model}
\begin{figure}[H]
    \centering
    \includegraphics[width=0.9\linewidth]{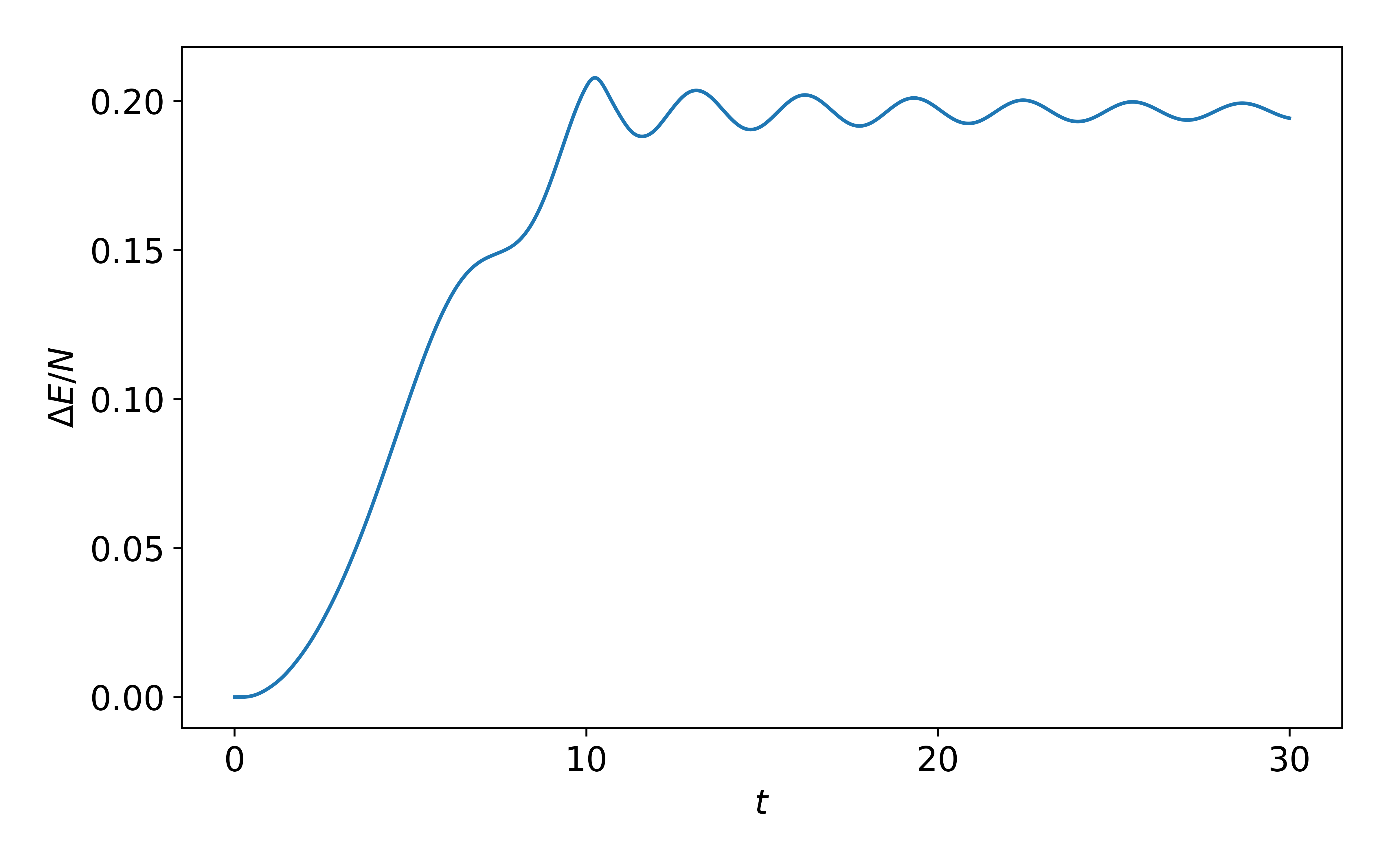}
    \caption{Energy per site (with $N = 300$) stored as function of time in an Ising-chain QB charged from an initial transverse field $h_i = 0.8$ to a final value $h_f = 1.5$ by means of a finite-time ramp with $t_f = 10$ (quench rate $v=0.07$).}
    \label{Energy_noiseless_ising}
\end{figure}

We now return to the quantum Ising chain in Sec 2.1, with the first-quantized mode Hamiltonians $H_{0,k}(t)$ in Eq. \eqref{Hamiltonian_Per_Mode}. Analogous to the single-qubit case in the previous section, we define
a single-mode Ising QB Hamiltonian
\begin{equation} \label{IsingmodeQB}
H_{\text{Batt},k} \equiv H_{0,k}(0).
\end{equation}
With this, we perform a numerical analysis that consists in computing the energy stored for each mode with respect to the initial ground‐state energy
\begin{equation} \label{Energy_Stored_Per_Mode}
\Delta E_k(t) = \mbox{Tr}[(\rho_{0,k}(t) - \rho_{0,k}(0)) \,H_{\text{Batt},k}],
\end{equation}
summing then for all modes \(k\) and finally dividing by the number of sites $N$ to obtain the stored energy per site $\Delta E / N$. From the plot reported in Fig. \ref{Energy_noiseless_ising}, choosing $N = 300$ and the same values of the parameters $h_i$ and $h_f$ as in Fig. \ref{DeltaESingoloQubitPlot} (for comparison with the single-qubit QB), we can see that considering a spin chain instead of a single-qubit system gives two main advantages: the first one is quantitative since the quantum Ising chain is able to store more energy per site, while the second one is qualitative and it is related to the smaller oscillations appearing both on the ramp part and on the plateau once $h_f$ is reached. This behavior can be understood by considering that in the quantum Ising chain the total stored energy is a sum over independent $k$-modes, each oscillating at its own well-defined frequency. Although each mode retains phase coherence and continues to oscillate indefinitely, the superposition of all modes leads to a dephasing effect, suppressing the clean, single-frequency oscillations observed in the single-qubit case and resulting in a smoother time evolution of the total stored energy.
Moreover, the ramp introduces a controllable time-scale for the charging that is absent in the sudden quench case for an Ising QB, where the short time behavior is dominated by strong and peaked oscillations \cite{Grazi24}.
\subsection{Noisy charging of the quantum Ising model}
\begin{figure}[H]
\centering

\begin{subfigure}{0.6\textwidth}
    \centering
    \includegraphics[width=\textwidth]{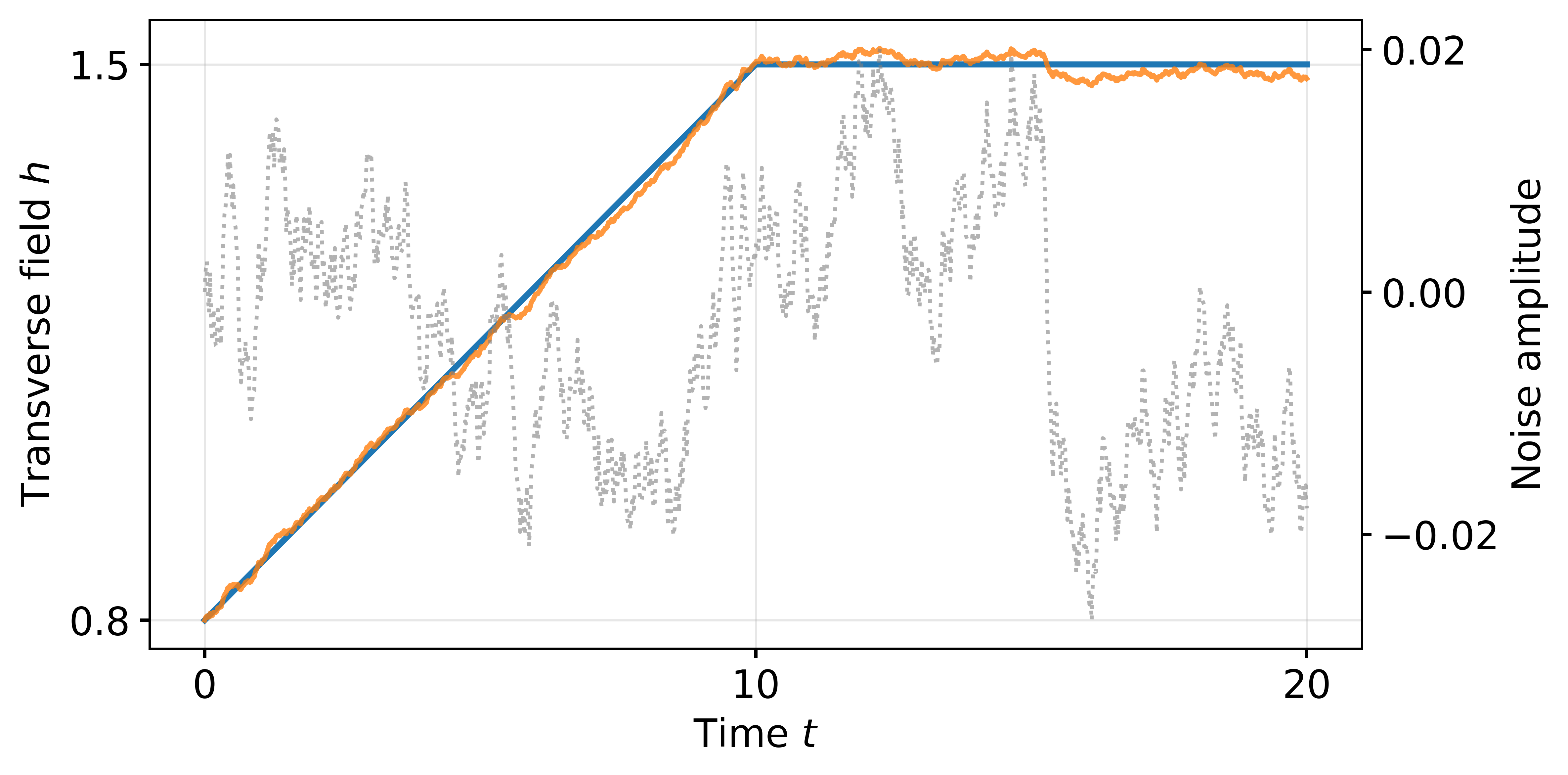}
    \caption{} 
\end{subfigure}
\hspace{0.05\textwidth}
\begin{subfigure}{0.6\textwidth}
    \centering
    \includegraphics[width=\textwidth]{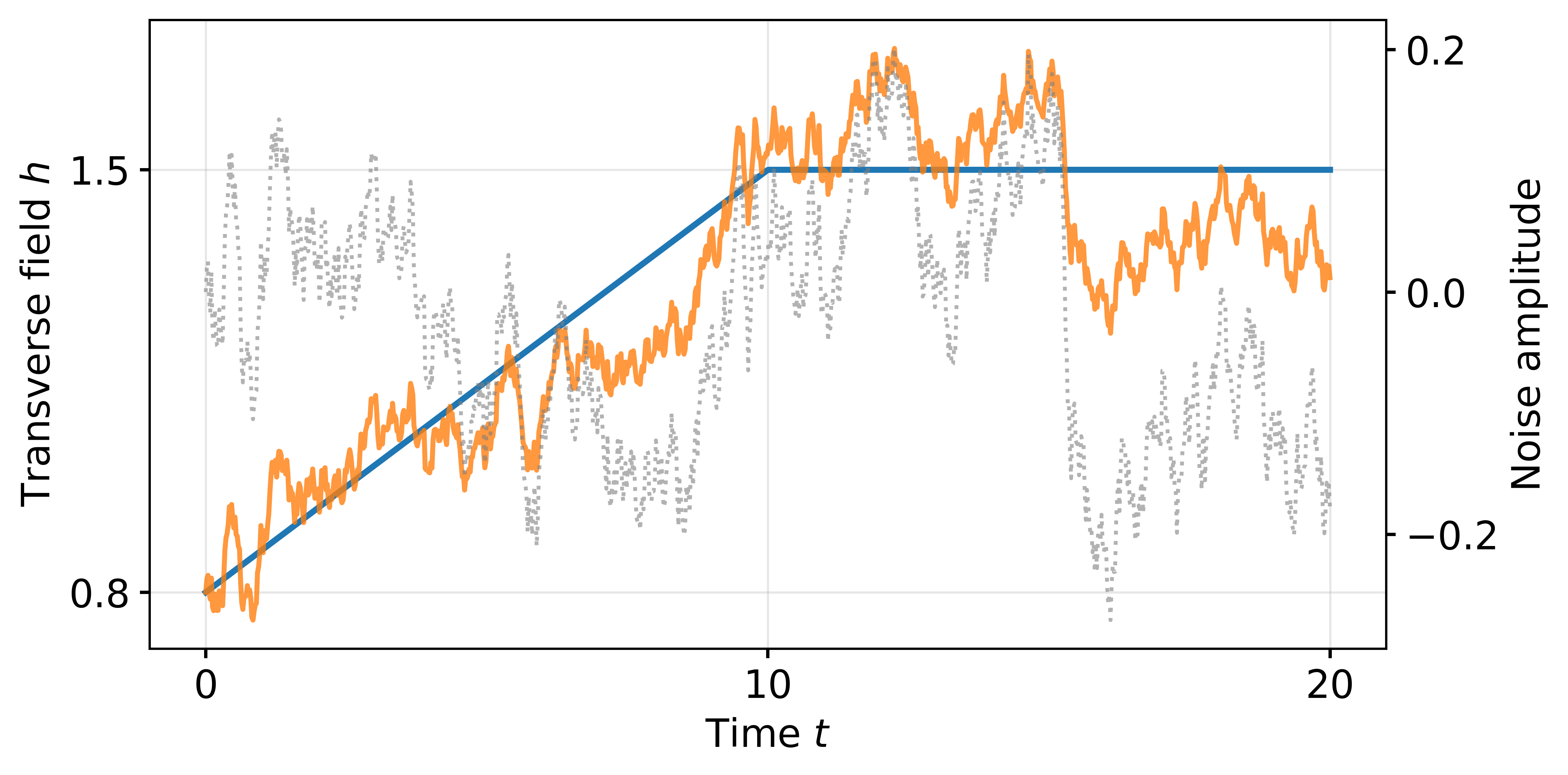}
    \caption{} 
\end{subfigure}

\vspace{0.8em}
\begin{subfigure}{0.6\textwidth}
    \centering
    \includegraphics[width=\textwidth]{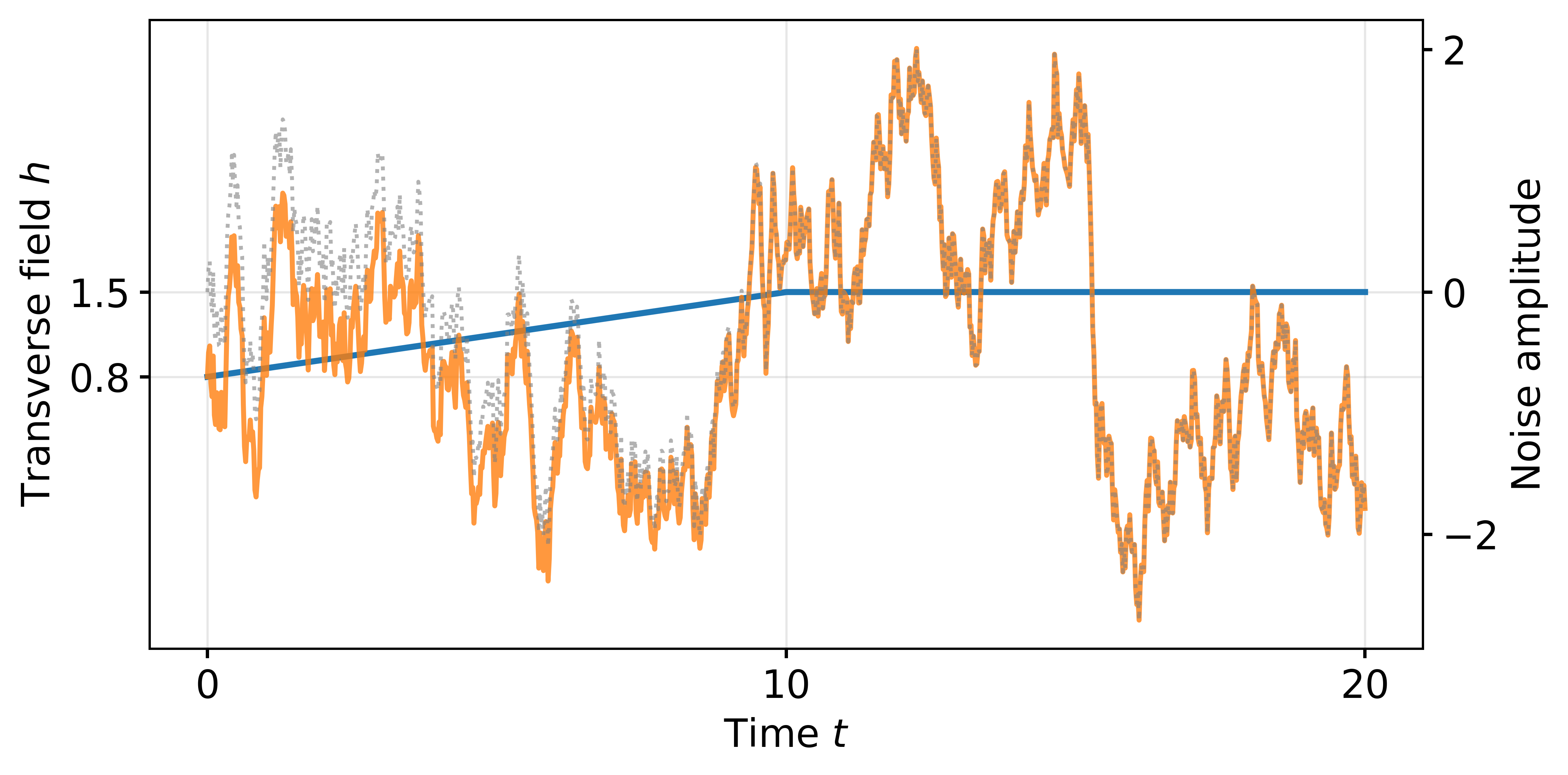}
    \caption{} 
\end{subfigure}

\caption{Charging protocol including noise with (a) $\xi = 0.01$ (b) $\xi = 0.1$ and (c) $\xi = 1$. In each panel the blue line shows the noiseless protocol in Eq. \eqref{ramp-quench}, the orange graph includes the effect of a single realization of Ornstein-Uhlenbeck noise with $\tau_n =1$ (obtained from numerical simulations on Python), and the gray dashed graph displays the noise signal alone, whose amplitude is reported on the right axis.}
\label{noisy_ramps}
\end{figure}

We now aim at addressing the robustness of the spin chain QB discussed above with respect to noise added to both segments of the charging protocol, and, more generally, the effect of noise. The protocol will now be modeled by replacing $h(t)$ by $h(t) + \eta(t)$, where $\eta(t)$ is a Gaussian distributed stochastic variable with zero mean, \(\langle \eta(t) \rangle = 0\), and Ornstein-Uhlenbeck correlations \cite{Jafari24, Wilkinson10, Lehle18, Maller09}
\begin{equation}
\bigl\langle \eta(t)\eta(t') \bigr\rangle = \frac{\xi^2}{2\tau_n} \exp\left(-\frac{|t - t'|}{\tau_n}\right),
\label{noise_corr}
\end{equation}
where $\xi$ is the noise intensity and $\tau_n$ its correlation time (from now on we will set  $\tau_n = 1$ since, as seen in our numerics, quantities of interest exhibit only a rather weak dependence on the value of $\tau_n$). Ornstein-Uhlenbeck noise serves as a paradigm for time-correlated noise, also known as ``colored noise", ubiquitous in any process (like the time-dependent ramp in the present problem) influenced by random motion of particles. The more commonly used, but idealized, limit of (correlated) white noise is obtained by taking $\tau_n \rightarrow 0$ in Eq. \eqref{noise_corr}. The noisy charging protocol is illustrated in Fig. \ref{noisy_ramps} for three different values of noise intensity, $\xi = 0.01, 0.1, 1$. As expected, for a sufficiently large noise intensity (here, $\xi=1$), there is no signature of the ramp anymore.

Recalling the Hamiltonian for each quasimomentum mode, reported in Eq. \eqref{Hamiltonian_Per_Mode}, our noisy ramp modifies $H_{0,k}(t)$ into
\begin{equation}
H_{k}(t) = H_{0,k}(t) + \eta(t)\,H_1, \label{Mode_Ham_With_Noise}
\end{equation}
where 
\begin{equation}
  H_{1} =  \begin{pmatrix}
       2 & 0 \\
        0 & -2
    \end{pmatrix}.
\end{equation}
In order to provide a more complete characterization of the Ising QB when charged by a noisy protocol, we now introduce another figure of merit alongside the stored energy: the ergotropy~\cite{Allahverdyan04}. This quantity is defined as the maximum amount of work that can be extracted from a quantum state by means of unitary transformations, bringing the system to a passive state, a state from which no further work can be extracted unitarily~\cite{Pusz78, Lenard78}. Given a density matrix $\rho$ describing a quantum state, we define as ergotropy the quantity
\begin{equation}
    \mathcal{E}(\rho) = \text{Tr}\{\rho H\} - \min_U \text{Tr}\{U\rho U^\dagger H\}, \label{Ergotropy_Definition}
\end{equation}
where the minimum is taken over all unitary operations \(U\), and the final term corresponds to the energy of the passive state associated with \(\rho\), namely the state that has the same spectrum as $\rho$ but is diagonal in the energy eigenbasis of $H$, with the eigenvalues of $\rho$ rearranged in decreasing order and assigned to the energy levels in increasing order. By construction, no work can be extracted by the passive state in an unitary way. The reason why we did not consider ergotropy in the noiseless scenario is that in absence of noise the ramp charging protocol is unitary: as a consequence, all the energy stored in the system remains extractable, meaning that the ergotropy is equal to the stored energy at all times: \(\mathcal{E}(t) = \Delta E(t)\). However, once the noise is introduced, the dynamics becomes non-unitary after averaging over single noise realizations, in full analogy with what happens due to the coupling of a QB with an effective dissipative environment~\cite{Breuer07, Weiss12, Carrega16, Farina19, DeFilippis23}. One therefore expects that part of the injected energy becomes non-extractable, which means it contributes to the stored energy but not to the ergotropy. To quantify this effect, we define a time-dependent efficiency of the QB as
\begin{equation}
    \epsilon(t) = \frac{\mathcal{E}(t)}{\Delta E(t)}, \label{DEF_Efficiency}
\end{equation}
which satisfies \(\epsilon(t) \leq 1\) at all times, with the equality holding in the noiseless case.

Considering our noisy charging protocol, with mode Hamiltonians as defined in Eq. \eqref{Mode_Ham_With_Noise}, we are interested in the expected performance of the Ising QB obtained by averaging over the Ornstein-Uhlenbeck noise distribution $\{ \eta(t) \}$. Thus, we shall consider the ensemble-averaged mode density matrix $\rho_k(t) \equiv \langle \rho_{\eta, k}(t) \rangle $ with $\langle \ldots \rangle$ the ensemble average over single noise realizations. As shown in Ref. \cite{Jafari24}, $\rho_k(t)$ can be obtained by numerically solving the coupled differential equations
\begin{equation}
    \begin{cases}
        \dot{\rho}_k(t) = -i[H_{0,k}(t), \rho_k(t)] - \frac{\xi^2}{2 \tau_n} [H_1, \Gamma_k(t)]\\[8pt]
        \dot{\Gamma}_k(t) = -\frac{\Gamma_k(t)}{\tau_n} + [H_1, \rho_k(t)]
    \end{cases}
    \label{diff_eq}
\end{equation}
with
\begin{equation}
    \Gamma_k(t) = \int_{t_i}^{t} e^{-(t - s)/\tau_n} [H_1, \rho_k(s)]\, ds,
\end{equation}
resembling a non-Markovian kernel for dissipative master equations~\cite{Breuer07, Breuer16}. Since our spin-chain quantum battery, before starting the charging protocol, is a closed quantum system initialized in a noiseless ground state, we impose the initial conditions $\rho_k(t_i) = \ket{\chi^-_k(t_i)}\bra{\chi^-_k(t_i)}$ and $\Gamma_k(t_i) = 0$ for each mode $k$, with the initial conditions on the second segment of the charging protocol determined by the final values of $\rho_k(t)$ and $\Gamma_k(t)$ of the preceding segment. Now we carry out the numerics, use Eq. \eqref{Energy_Stored_Per_Mode} and sum over all modes; this yields the total stored energy $\Delta E$, and via Eq. \eqref{Ergotropy_Definition}, the ergotropy $\mathcal{E}(t)$ as a function of time for different amplitudes $\xi$ of the noise. 

\begin{figure}[H]
\centering

\begin{subfigure}{0.6\textwidth}
    \centering
    \includegraphics[width=\textwidth]{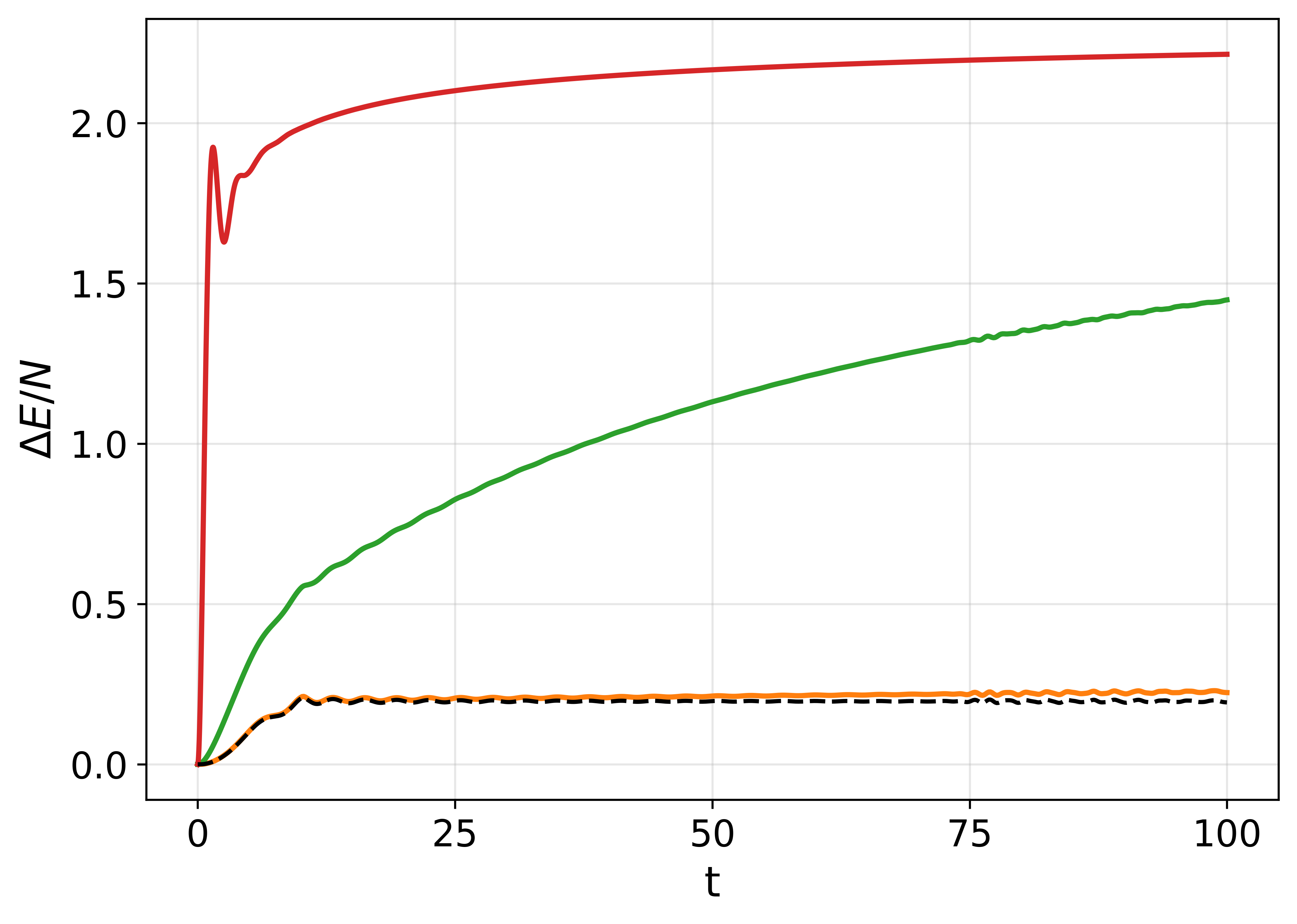}
    \caption{} 
    \label{Energy_Ramp}
\end{subfigure}
\hspace{0.05\textwidth}
\begin{subfigure}{0.6\textwidth}
    \centering
    \includegraphics[width=\textwidth]{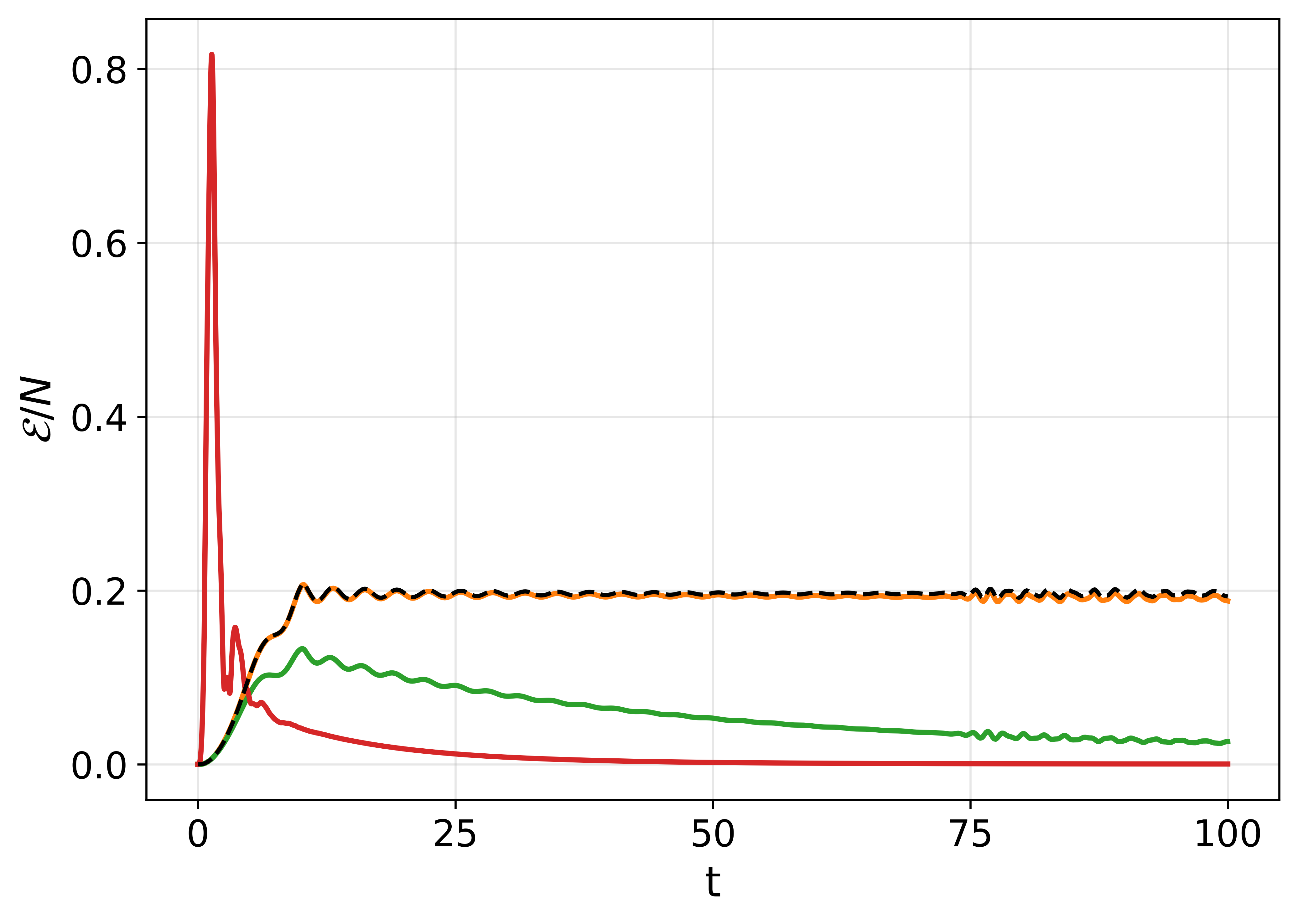}
    \caption{} 
    \label{Ergotropy_Ramp}
\end{subfigure}
\hspace{0.05\textwidth}
\begin{subfigure}{0.6\textwidth}
    \centering
    \includegraphics[width=\textwidth]{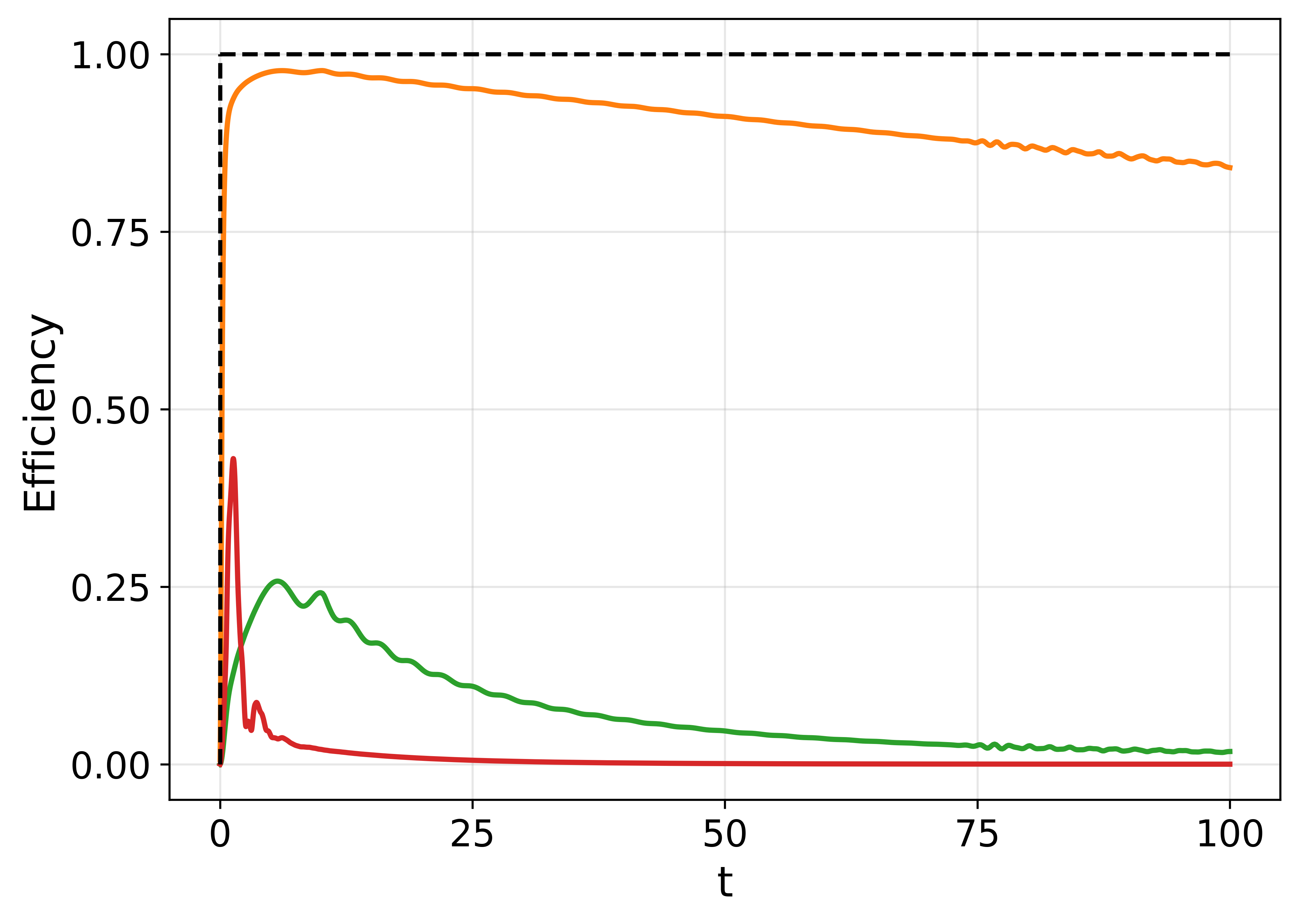}
    \caption{} 
    \label{Efficiency_Ramp}
\end{subfigure}

\caption{(a) Stored energy per site, (b) ergotropy per site and (c) efficiency of the Ising quantum battery as functions of time for the noiseless case $\xi = 0$ (dashed black) and for noise intensities $\xi = 0.01$ (orange), $\xi = 0.1$ (green) and $\xi = 1$ (red) added to the charging protocol in Eq. \eqref{ramp-quench} from $h_i = h(t = 0) = 0.8$ to $h_f = h(t = 10) = 1.5$ (quench rate $v=0.07$).}
\label{Energy_Ergotropy_Efficiency_Ramp}
\end{figure}

Results for the stored energy per site $\Delta E(t)/N$, ergotropy per site $\mathcal{E}(t)/N$, and efficiency $\epsilon(t)$ are plotted in Fig. \ref{Energy_Ergotropy_Efficiency_Ramp} for different choices of noise intensity $\xi$. From Fig. \ref{Energy_Ramp} we can observe that, as the noise intensity increases, more non-extractable energy is injected into the system: in particular, when the intensity becomes comparable to the jump between $h_i$ and $h_f$, the effect of the ramp protocol becomes negligible compared to the environmental noise and the characteristic features of the underlying noiseless charging process are completely lost. The abrupt energy injection right at the start of the ramp produces, in the $\xi = 1$ case, a pronounced peak of ergotropy close to $t = 0$, as shown in Fig. \ref{Ergotropy_Ramp}, while for small noise intensities the ergotropy remains almost identical to the noiseless case, resulting in an efficiency above $80~\%$ for $\xi = 0.01$. At fixed times after the ramp, efficiency, Fig. \ref{Efficiency_Ramp}, decays exponentially with noise intensity. The abrupt energy injection when $\xi=1$ reflects that for larger noise intensities, quasiparticles are easily excited across the energy gap, implying a very fast increase of the energy.

We stress the fact that the choice of ramp values $(h_i = 0.8,\, h_f = 1.5)$ is arbitrary and that crossing the quantum critical point at $h=1$ is not essential for the phenomena discussed here, as the conclusions we draw do not rely on this particular change of quantum phase from ferromagnetic to paramagnetic. Similarly to previous works \cite{Grazi24, Grazi25, Grazi25Energies},  the only situation in which the proximity to a critical point becomes operationally relevant is when one aims to maximize the total stored energy: in that case, the most favorable choice is to quench to a final field $h_f$ as close as possible to the critical values. 

\begin{figure}[H]
\centering

\begin{subfigure}{0.6\textwidth}
    \centering
    \includegraphics[width=\textwidth]{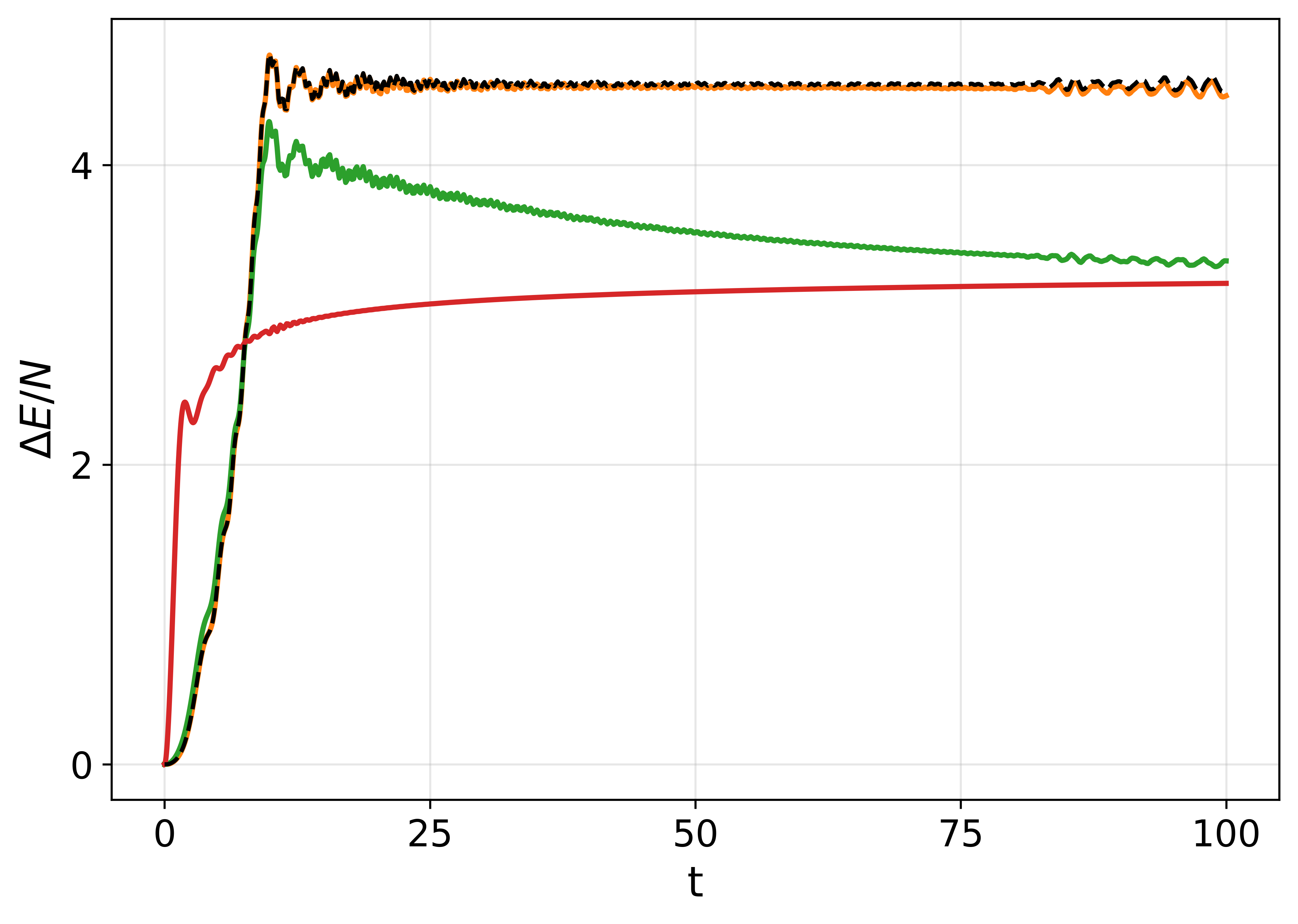}
    \caption{} 
    \label{Energy_DoublePhaseRamp}
\end{subfigure}
\hspace{0.05\textwidth}
\begin{subfigure}{0.6\textwidth}
    \centering
    \includegraphics[width=\textwidth]{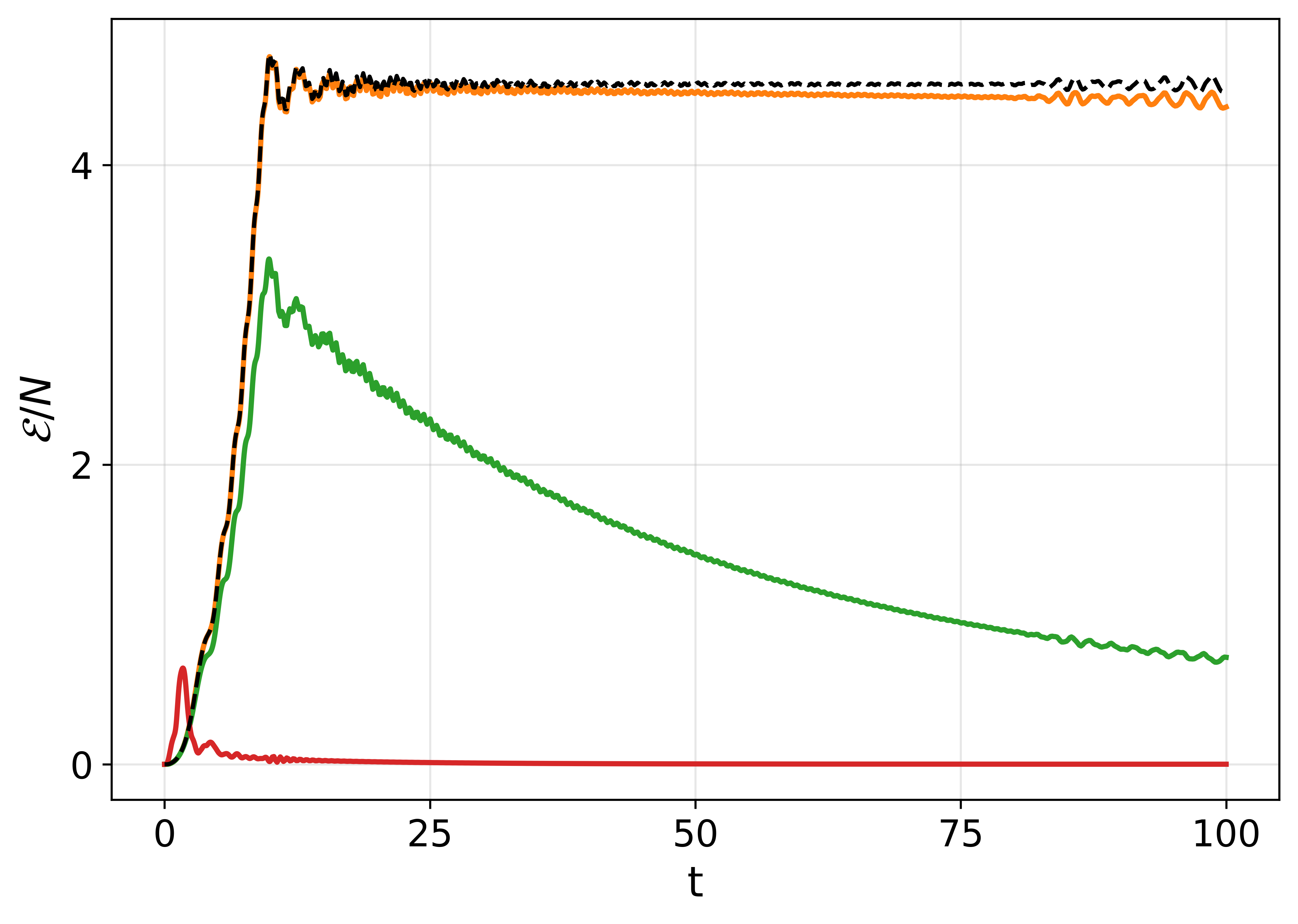}
    \caption{} 
    \label{Ergotropy_DoublePhaseRamp}
\end{subfigure}
\hspace{0.05\textwidth}
\begin{subfigure}{0.6\textwidth}
    \centering
    \includegraphics[width=\textwidth]{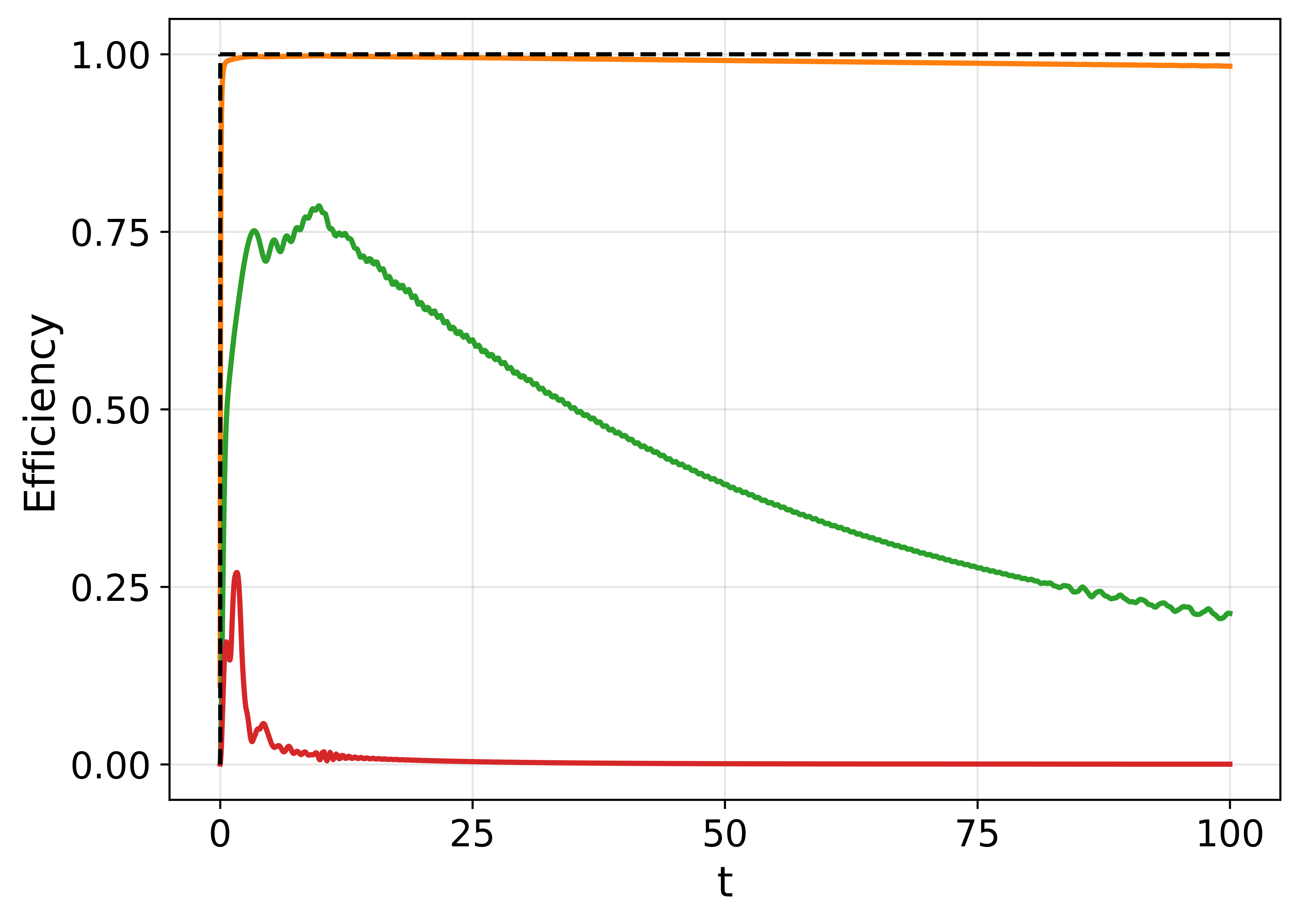}
    \caption{} 
    \label{Efficiency_DoublePhaseRamp}
\end{subfigure}

\caption{(a) Stored energy per site, (b) ergotropy per site and (c) efficiency of the Ising quantum battery as functions of time for the noiseless case $\xi = 0$ (dashed black) and for noise intensities $\xi = 0.01$ (orange), $\xi = 0.1$ (green) and $\xi = 1$ (red) added to the charging protocol in Eq. \eqref{ramp-quench} from $h_i = h(t = 0) = -1.5$ to $h_f = h(t = 10) = 1.5$ (with quench rate $v=0.3$).}
\label{Energy_Ergotropy_Efficiency_DoublePhaseRamp}
\end{figure}

Interestingly, we find that the enhancement of stored energy with noise intensity is not a universal behavior. In fact, if we consider a ramp that goes from $h_i = -1.5$ to $h_f = 1.5$, a qualitatively different behavior emerges, as illustrated in the three panels of Fig. \ref{Energy_Ergotropy_Efficiency_DoublePhaseRamp}. In this scenario, larger noise intensities correspond to lower stored energy, while the robustness improves. For instance, at $\xi = 0.01$ the efficiency remains close to $100~\%$ and even for intermediate noise values, the performances improve significantly, with the efficiency increasing from a maximum of about $25~\%$ to approximately $75~\%$.

\begin{figure}[H]
\centering

\begin{subfigure}{0.6\textwidth}
    \centering
    \includegraphics[width=\textwidth]{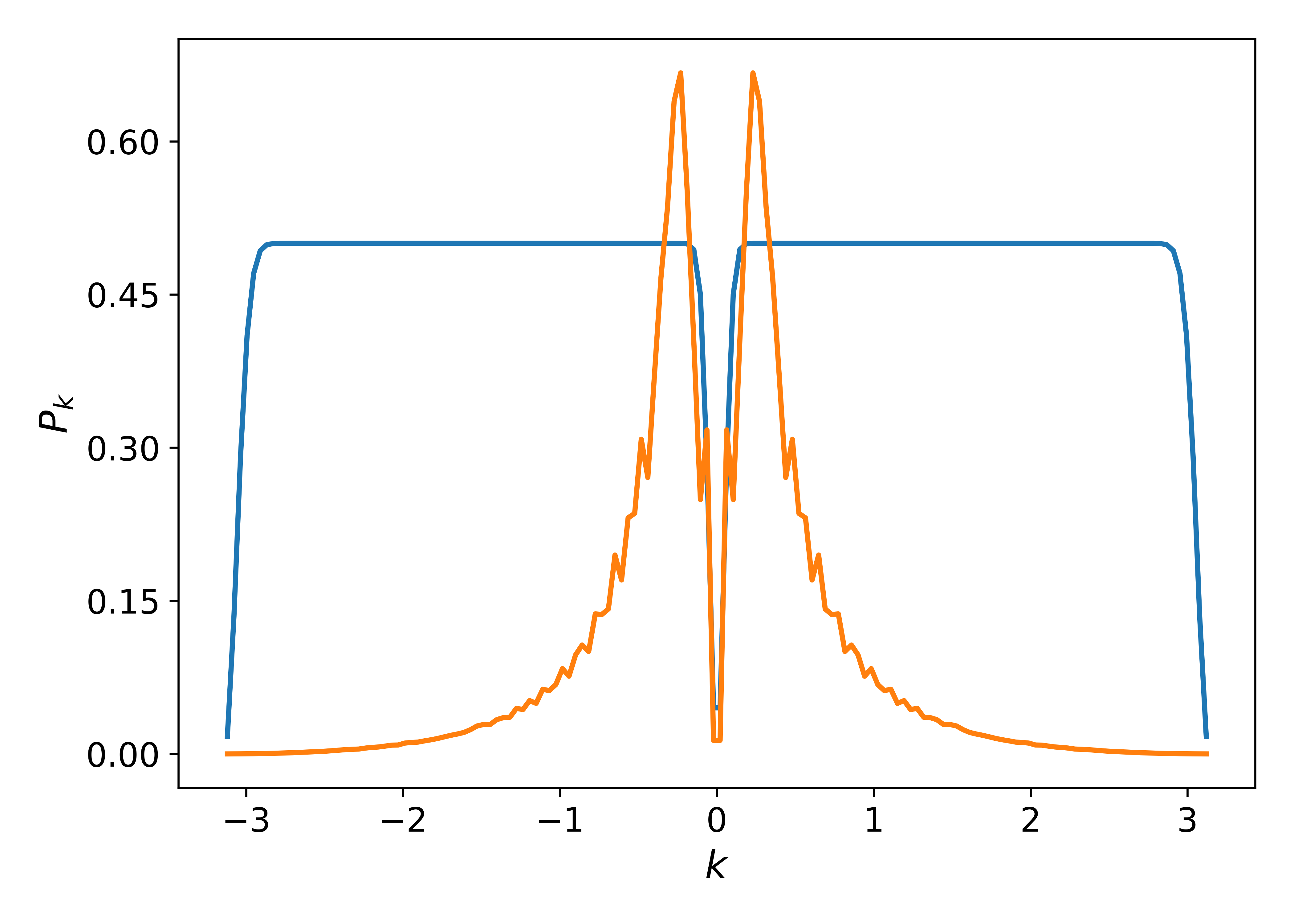}
    \caption{} 
    \label{Pk_08}
\end{subfigure}
\hspace{0.05\textwidth}
\begin{subfigure}{0.6\textwidth}
    \centering
    \includegraphics[width=\textwidth]{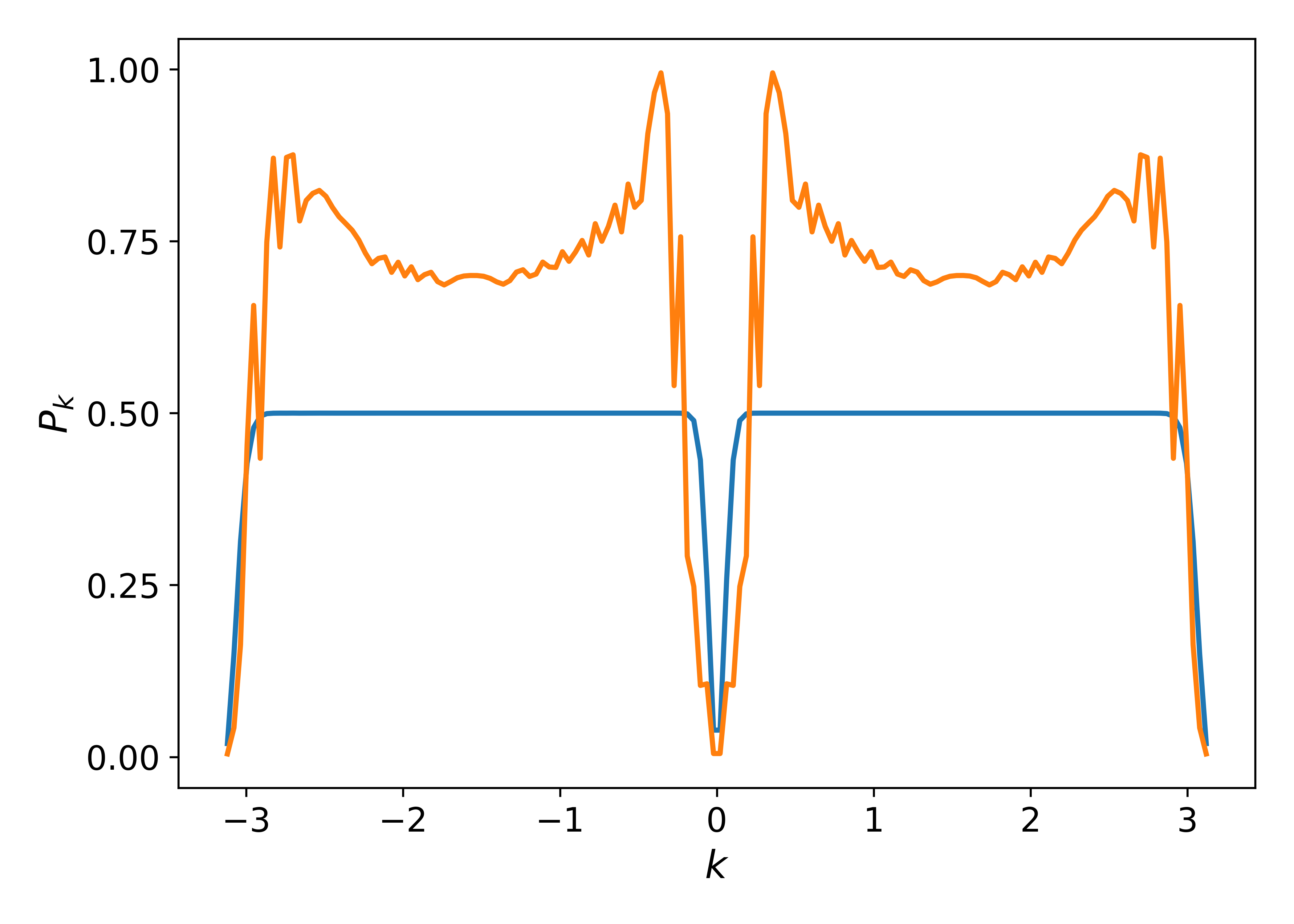}
    \caption{} 
    \label{Pk_pm}
\end{subfigure}

\caption{Mode-resolved excitation probability $P_k$ for the quench (a) from $h_i = 0.8$ to $h_f = 1.5$ and (b) from $h_i = -1.5$ to $h_f = 1.5$. The orange curve represents the noiseless ramp, while the blue curve represents the ensemble-averaged noisy ramp having intensity $\xi = 1$.}
\label{Mode_resolved_probability}
\end{figure}

To identify the origin of this behavior, we analyzed different quenches and found that the key quantity controlling whether strong noise enhances or suppresses stored energy is the mode-resolved excitation probability at the end of the protocol
\begin{equation}
    P_k = \bra{\chi_k^+(t^*)}\rho_k(t^*)\ket{\chi_k^+(t^*)}.
\end{equation}
where $t^* > t_f$ is the time we disconnect the battery from the charger, i.e. $h(t^*) = h_i = h(0)$. By comparing the noiseless and most noisy cases (here $\xi = 1$) for both ramps, we observe that the noisy profiles are essentially identical: noise flattens the excitation probability to $P_k = 0.5$ for all modes except $k=0,\pm\pi$. In contrast, the noiseless behavior is very different, since for the quench starting at $h_i = 0.8$ (Fig.~\ref{Pk_08}), only a few modes exhibit $P_k > 0.5$, whereas for the quench starting at $h_i=-1.5$ (Fig.~\ref{Pk_pm}), nearly all modes satisfy $P_k > 0.5$. This observation explains the trends shown in Fig. \ref{Energy_Ramp} and Fig. \ref{Energy_DoublePhaseRamp} as follows. When the noise intensity $\xi$ is larger than the quench rate $v=\Delta h/\Delta t$ ($\xi/v \approx 14$ in Fig. \ref{Energy_Ramp} and $\xi/v \approx 3$ in Fig. \ref{Energy_DoublePhaseRamp}), ensemble-averaged noise tends to push most of the excitation probabilities towards the value $P_k=0.5$, signaling the appearance of a maximally mixed state for the corresponding modes \cite{Jafari24}.

If the noiseless protocol excites only a small subset of modes above this threshold, noise raises their population and therefore increases the stored energy. Otherwise, if the noiseless dynamics already produces $P_k>0.5$ for a large fraction of the spectrum, noise drives those modes downward toward $0.5$, resulting in a net suppression of the stored energy. We recall that, as shown in Fig.~\ref{Efficiency_Ramp} and Fig.~\ref{Efficiency_DoublePhaseRamp}, a larger amount of stored energy does not necessarily imply better performance, since the relevant operational quantity in this sense is the ergotropy. As discussed above, ergotropy depends not only on how much energy is stored but, more importantly, on how that energy is distributed among the eigenmodes. For this reason, the most efficient protocol is the one shown in Fig.~\ref{Efficiency_DoublePhaseRamp}.

\section{Discussion}
In this work, we have addressed strategies for the control of the stored energy and ergotropy in a transverse-field Ising chain considered as a quantum battery. By studying a charging process consisting of a finite-time ramp that goes from an initial field $h_i$ at time $t = 0$ to a final value $h_f$ at time $t = t_f$, we showed that, in a simple single-qubit system, the oscillations tend to follow the slope of the ramp and their amplitude decreases as the ramp duration increases (at fixed $h_i$ and $h_f$). Then, by addressing the Ising chain, we observed both a larger amount of stored energy per site and its more regular behavior with time both during and after the ramp when the field has taken a constant value $h_f$. Next, we studied the robustness of the stored energy and ergotropy when noise is superposed to the charging ramp for two different representative quenches, both ending at $h_f = 1.5$ but starting from $h_i = 0.8$ and $h_i = -1.5$, respectively. In the first case, we observed that by increasing the intensity of the noise, more energy is stored into the battery; however, most of this energy is non-extractable, since with this protocol noise degrades the ergotropy. In the second case, noise has the opposite effect, reducing the stored energy. However, this latter protocol turns out to be the more advantageous one from the point of view of quantum battery performance, since the efficiency, defined at each instant of time as the ratio between ergotropy and stored energy, remains higher than in the previous case, showing that a small amount of noise has almost no effect on the work extraction. In closing, given a noise intensity $\xi$ and a quench rate $v$ of a charging protocol, let us point out that the expected behavior under intermediate and strong noise with $\xi/v > 1$ is already
encoded in the noiseless ramp of the protocol: as implied by our discussion of the results displayed in Fig. \ref{Mode_resolved_probability}, it is sufficient to examine what proportion of the $k$ modes of the Ising Hamiltonian that are excited above $P_k = 1/2$.
This allows one to reliably predict whether noise will increase or decrease the energy stored in the battery.

\section{Supporting Material}
\subsection{Single-qubit charging}
Here, we consider in detail the effect of a finite-time charging ramp used to promote the simplest possible example of QB, namely a single qubit, from its ground to its excited state. The system is described by the Hamiltonian
\begin{equation}
    H(t) = -J\sigma^x - h(t) \sigma^z
\end{equation}
where $J$ provides the characteristic energy scale of the qubit in the absence of the driving field (in the main text we set $J = 1$), while $h(t)$ is the one reported in Eq. \eqref{ramp-quench}. As stated in the main text, $H_{B} \equiv H(t=0)$ indicates the free Hamiltonian of the QB, while the remaining part is associated with the time-dependent action of an external classical charger. For generic values of $J$, the normalized eigenstates of $H_B$ are
\begin{equation}
    \ket{\pm} \equiv \frac{1}{\sqrt{2\omega_B(\omega_B\mp h_i)}} \begin{pmatrix}
     h_i \mp  \omega_B   \\   J
    \end{pmatrix} \label{Eigenstates_WITH_J}
\end{equation}
having energy $\pm \omega_B$ respectively, with $\omega_B \equiv \sqrt{J^{2}+h_i^2}$, while
\begin{equation}
    E(t)=\text{Tr}[\rho(t)H_{B}] = -h_i\left(\rho_{00}(t) - \rho_{11}(t)\right) - J\left(\rho_{01}(t) +\rho_{10}(t)\right),
\end{equation}
which leads to the stored energy reported in Eq. \eqref{stored}.
As in the main text, we consider $\ket{-}$ as the initial state of the QB and from this state we obtain the initial density matrix
\begin{equation}
    \rho(0) = \ket{-}\bra{-} = \frac{1}{2\omega_B(\omega_B+h_i)} \begin{pmatrix}
       (\omega_B + h_i)^2  & J\left(\omega_B + h_i\right)\\ J\left(\omega_B + h_i\right) & J^{2}
    \end{pmatrix}. \label{rho_zero_WITH_J}
\end{equation}
After repeating the mathematical steps of the main text and introducing the set of functions reported in Eq. \eqref{FuncsUXY}, we need to solve the following system of differential equations
\begin{equation}
    \begin{cases}
        &\dot{u} = 2Jy\\
        &\dot{x} = -2h(t)y\\
        &\dot{y} = -2Ju + 2h(t)x
    \end{cases} \label{New_Eqs}
\end{equation}
in order to compute the stored energy
\begin{equation}
     \Delta E(t) = -h_{i}\left[u(t)-u(0) \right]-J\left[x(t)-x(0) \right]. \label{Energy_u_x}
\end{equation}
To get an intuition about the time evolution governed by this system of equations we start by considering $h(t)$ as constant in time, $h(t) = h_0$. In this case one can differentiate the third equation in the system above and then use the first two equations to substitute for $\dot{u}$ and $\dot{x}$. This yields to
    \begin{equation}
        \ddot{y} = -2J\dot{u} + 2h_0\dot{x} = -4(J^{2} + h_0^2) y.
    \end{equation}
Therefore $y(t)$ satisfies the differential equation of a harmonic oscillator with angular frequency $\Omega = 2\sqrt{J^{2} + h_0^2}$, which notably equals the energy gap between the ground state $\ket{-}$ and the excited state of the qubit $\ket{+}$ when $h_0 = h_i$ (cf. Eq. \eqref{Eigenstates_WITH_J}). A consequence of this fact is that the components of $\rho(t)$ and the stored energy will show an oscillating behavior in the regions where the external charging field is costant (in our case, at the end of the ramp). Taking into account the time dependence of $h(t)$, the same steps lead to an inhomogeneous forced oscillator equation for \(y(t)\), namely 
\begin{equation}
\ddot y + 4\big(J^{2}+h(t)^2\big)\,y \;=\; 2\dot h(t)\,x. 
\label{Eq_Diff_Forced}
\end{equation}
In other words, the system behaves as an oscillator with instantaneous angular frequency
\begin{equation}
\Omega(t)=2\sqrt{J^{2}+h(t)^2}, \label{Omega}
\end{equation}
driven by a source proportional to the product of $\dot{h}(t)$ and $x(t)$ (with $\dot{h}(t) = v$ for the specific ramp considered in Eq. \eqref{ramp-quench}). Two regimes are particularly informative. In order to analyze them we will split the time-evolved density matrix in two parts:
\begin{equation}
    \rho(t) = \rho_{slow}(t) + \delta \rho(t)
\end{equation}
where $\rho_{slow}(t)$ is the slowly varying adiabatic approximation, while $\delta \rho(t)$ contains nonadiabatic oscillatory corrections. Inserting this form into Eq. \eqref{stored} one obtains
\begin{equation}
    \Delta E(t) = \Delta E_{slow}(t) + \Delta E_{osc}(t)
\end{equation}
where $\Delta E_{slow}(t) = \text{Tr}[(\rho_{slow}(t) - \rho(0)) H_B]$ and $\Delta E_{osc}(t) = \text{Tr}[\delta \rho(t) H_B]$. Let's focus on the ``slow" part first. If the quench rate $v$ is small compared with the energy gap $2\omega_B$, the state remains close to the instantaneous ground state, hence the density matrix of the state $\rho(t)$ can be written as the one for the ground state in Eq. \eqref{rho_zero_WITH_J} where $h_{i}$ is replaced with $h(t)$. This leads to
\begin{equation}
u_{\mathrm{slow}}(t) \approx \frac{h(t)}{\sqrt{J^{2}+h(t)^2}},\qquad
x_{\mathrm{slow}}(t) \approx \frac{J}{\sqrt{J^{2}+h(t)^2}},\qquad
y_{\mathrm{slow}} \approx 0,
\end{equation}
so that $\Delta E_{slow}$ can be obtained by substituting these expressions into Eq. \eqref{Energy_u_x}, obtaining
\begin{eqnarray}
    \Delta E_{slow}(t) &\approx& -h_{i}\left[u_{slow}(t)-u(0) \right]-J\left[x_{slow}(t)-x(0) \right]\nonumber\\
    &\approx&-h_i\left(\frac{h(t)}{\sqrt{J^{2}+h(t)^2}} - \frac{h_i}{\sqrt{J^{2}+h_i^2}}\right)-J^{2}\left(\frac{1}{\sqrt{J^{2}+h(t)^2}}-\frac{1}{\sqrt{J^{2}+h_i^2}}\right).\nonumber\\ \label{Eslow}
\end{eqnarray}
If we now consider the nonadiabatic corrections, we can write $\Delta E_{osc}(t)$, in analogy with Eq. \eqref{Energy_u_x}, in the form
\begin{equation}
    \Delta E_{osc}(t) = -J x_{osc}(t) - h_i u_{osc}(t). \label{E_osc}
\end{equation}
To find the new functions $x_{osc}(t)$ and $u_{osc}(t)$, one can linearize the forcing term by replacing \(x(t)\) in Eq. \eqref{Eq_Diff_Forced} with \(x_{slow}(t)\), assuming that the corrections are small. This leads to an oscillating solution whose amplitude scales parametrically as
\begin{equation}
y_{osc}(t)\sim\frac{J\dot h(t)}{2\big(J^2+h(t)^2\big)^{3/2}}.
\label{approx_amp}
\end{equation}
We can now solve the system in Eq. \eqref{New_Eqs} for the $\alpha_{osc}(t)$ functions ($\alpha = u,x,y$) and substitute them in Eq. \eqref{E_osc}. This will lead to an oscillating correction to the stored energy with amplitude
\begin{equation}
\Delta E_{\mathrm{osc,amp}}(t)\propto\frac{\dot h(t)}{\big(J^2+h(t)^2\big)^2}. \label{Energy_Osc_Amp}
\end{equation}
Thus, in this perturbative regime the oscillation amplitude is linear in the quench rate \(v\) and suppressed at large instantaneous gaps \(2\sqrt{J^2+h(t)^2}\). 

\begin{figure}[H]
    \centering
    \includegraphics[width=\linewidth]{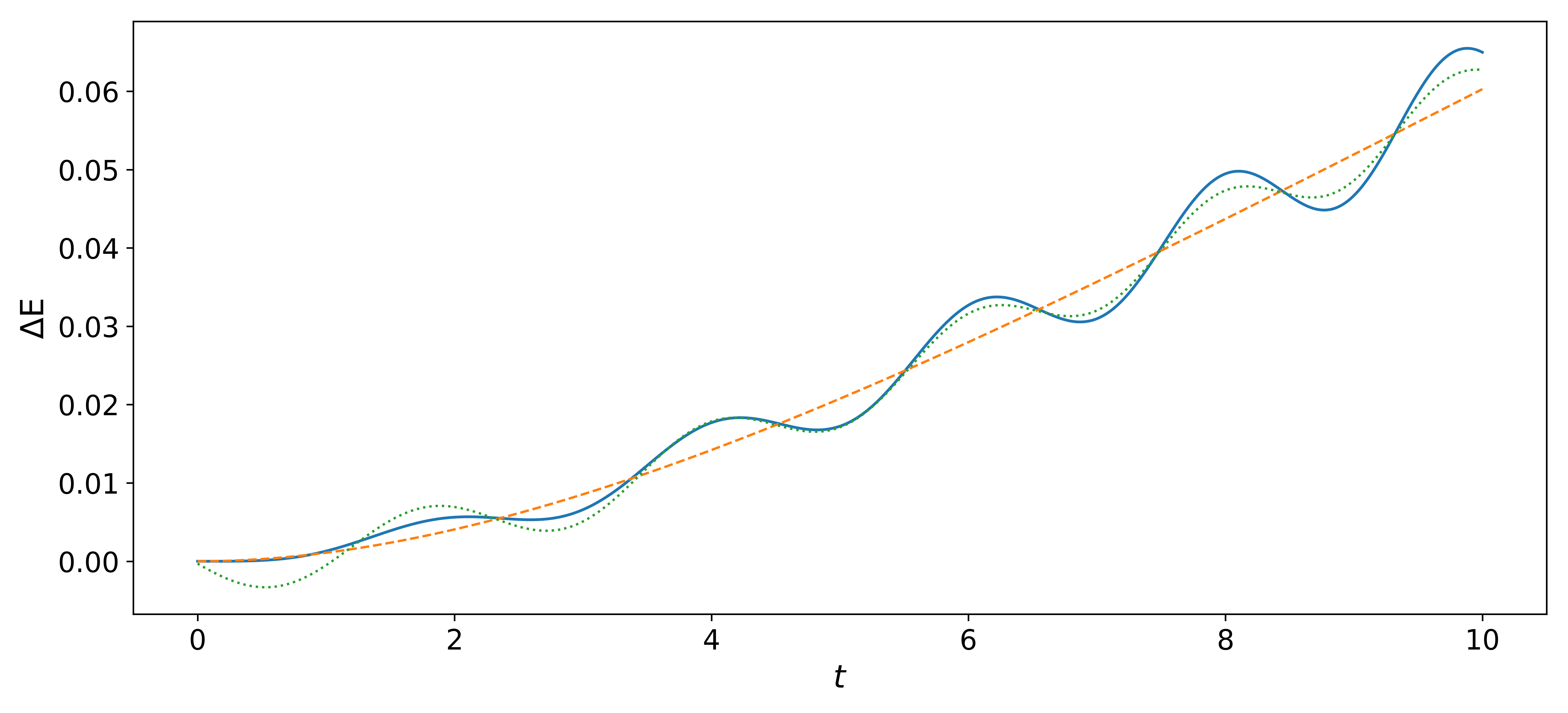}
    \caption{Energy stored during the ramp quench between $h_i = 0.8$ and $h_f = 1.5$ with $t_f = 10$ as a function of time, with quench rate $v=0.07$. The plot shows the numerically-computed stored energy (blue curve), the analytical $\Delta E_{slow}$ (orange curve) and the ansatz, described in the main text, that combines $\Delta E_{slow}$ and fitted oscillations (green curve).}
    \label{tr1_10}
\end{figure}

\begin{figure}[H]
    \centering
    \includegraphics[width=\linewidth]{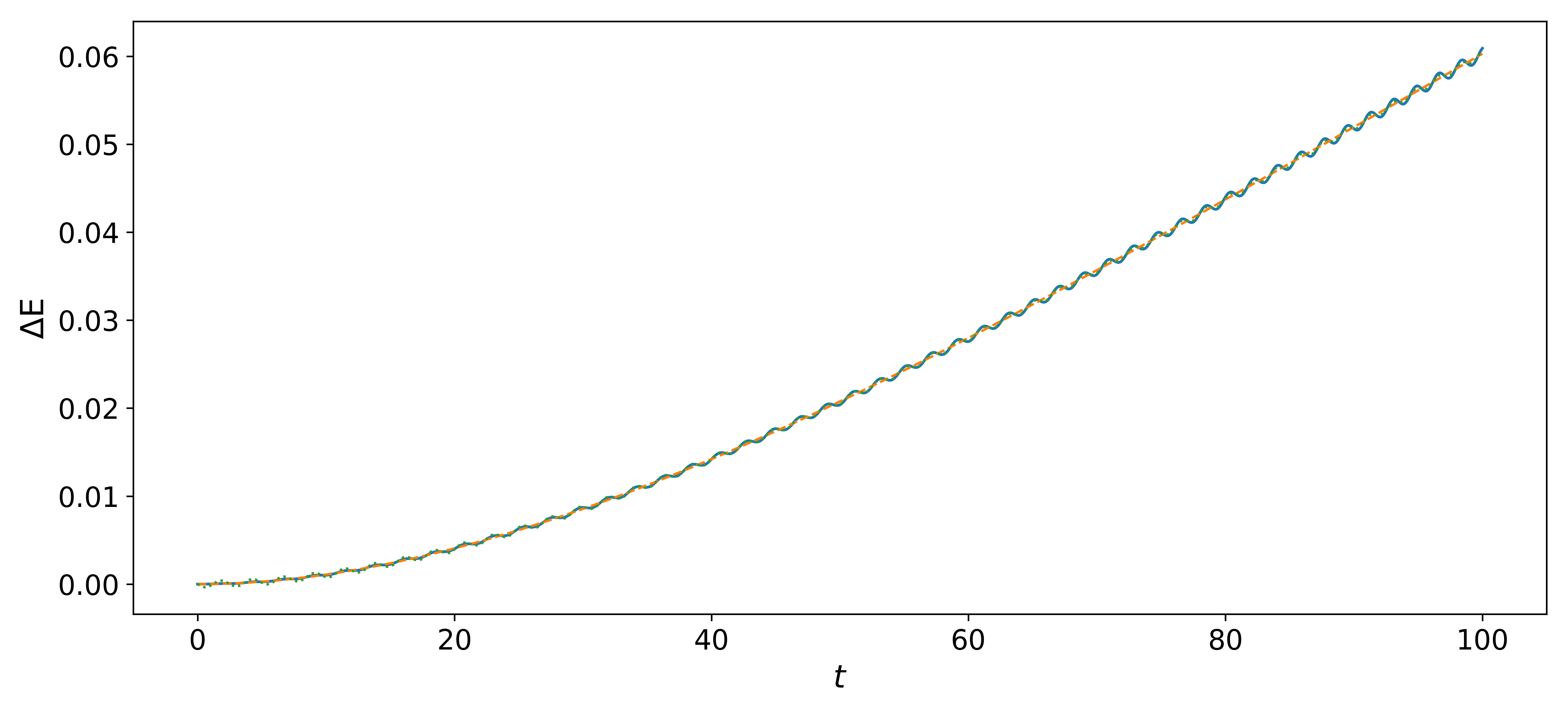}
    \caption{Energy stored during the ramp quench between $h_i = 0.8$ and $h_f = 1.5$ with $t_f = 100$ as a function of time, with quench rate $v=0.007$. The plot shows the numerically-computed stored energy (blue curve), the analytical $\Delta E_{slow}$ (orange curve) and the ansatz, described in the main text, that combines $\Delta E_{slow}$ and fitted oscillations (green curve).}
    \label{tr1_100}
\end{figure}

Motivated by these analytical results, we now turn to numerical simulations. In the following, we compare three curves: the full numerical solution for $\Delta E$ in presence of the linear quench, the analytical approximation $\Delta E_{\mathrm{slow}}(t)$ given in Eq. \eqref{Eslow} and an ansatz consisting of $\Delta E_{\mathrm{slow}}$ with superposed small oscillatory corrections of the form $A_{\mathrm{fit}}(t) \cos(\phi(t) - \phi_0)$ that mimic the oscillatory behavior given by $\Delta E_{osc}$(t). Here, $A_{\mathrm{fit}}(t)$ and $\phi_0$ are fitting parameters while
\begin{equation}
\phi(t)=\int_{0}^{t}\Omega(s)\,\mathrm{d}s,
\end{equation}
where $\Omega(s)$ is the frequency in Eq. \eqref{Omega}, equal to the instantaneous energy gap at time $s$. For sake of simplicity, as we did in the main text, in the following we will set the energy scale $J = 1$, so that all energies appearing in the plots will be in units of $J$ and time will have unit $J^{-1}$. We now analyze two different ramps: a faster quench from $t = 0$ to $t_{f} = 10$ (Fig. \ref{tr1_10}), and a slower one from $t = 0$ to $t_{f} = 100$ (Fig. \ref{tr1_100}), both with the same values of $h_i = 0.8$ and $h_f = 1.5$. These plots show two interesting features: first, we can observe that for faster ramps (Fig. \ref{tr1_10}), the approximation $\Delta E_{\mathrm{slow}}$ reproduces the dynamics accurately only at short enough times, after which oscillations dominate. Nevertheless, the  ansatz with oscillatory corrections superposed on $\Delta E_{\mathrm{slow}}(t)$ (green curve) still provides a quite good qualitative description of the numerical result. In contrast, and as expected, for slower ramps (Fig. \ref{tr1_100}), the agreement between $\Delta E_{\mathrm{slow}}$ and the numerical solution is very good across the full time evolution. The second observation is that the fitted amplitudes of the oscillatory corrections, which are
\begin{equation}
A_{\mathrm{fit}}(t_{f}=10) \approx 0.00366; \quad
A_{\mathrm{fit}}(t_{f}=100) \approx 0.000363,
\end{equation}
scale consistently with the linearization of Eq. \eqref{Energy_Osc_Amp}, showing that a faster ramp is subject to oscillations with a larger amplitude.

\noindent The previous observations are further strengthened by Fig. \ref{DeltaESingoloQubitPlot} of the main text, which shows an example of the complete evolution of $\Delta E(t)$ for a single qubit subject to the charging $h(t)$ in Eq. \eqref{ramp-quench}, i.e., including also the postquench dynamics with a constant transverse field $h_f$. Note the oscillations in $\Delta E(t)$ after the quench, predicted from Eq. \eqref{E_osc}.

\section{Methods}
Numerical simulations in the main text were performed in Python using NumPy, SciPy, and QuTiP. For each momentum sector, the time evolution of the density matrix was obtained through SciPy’s \textit{solve\_ivp} integrator with \textit{DOP853} and tolerances \textit{rtol = 1e-9} and \textit{atol = 1e-12}. Convergence with respect to time discretization and momentum resolution was verified. In the Supporting Materials, the Schrödinger equation for the two-level $k-$mode Hamiltonian was integrated using a standard fourth-order Runge-Kutta (RK4) scheme on a uniform time grid covering the ramp interval, while the optimal fitting parameters were obtained by linear least squares.

\section{Declarations}
\subsection{Resource Availability}
\subsubsection{Lead contact}
Requests for further information and resources should be directed to and will be fulfilled by the lead contact, Riccardo Grazi (riccardo.grazi@edu.unige.it).
\subsubsection{Materials availability}
This study did not generate new unique material.
\subsubsection{Data and code availability}
All original code used during the current study are available from the corresponding authors upon reasonable request.
\subsection{Acknowledgments}
{N.T.Z. acknowledges funding through the “Non-reciprocal supercurrent and topological transitions in hybrid Nb- InSb nanoflags” project (Prot. 2022PH852L) in the framework of PRIN 2022 initiative of the Italian Ministry of University (MUR) for the National Research Program (PNR). This project has been funded within the programme “PNRR Missione
4—Componente 2—Investimento 1.1 Fondo per il Programma Nazionale di Ricerca e Progetti
di Rilevante Interesse Nazionale (PRIN)”. D.F. and R.G. acknowledges funding from the European Union-
NextGenerationEU through the “Solid State Quantum Batteries: Characterization and Optimization” (SoS-QuBa) project (Prot. 2022XK5CPX), in the framework of the PRIN 2022 initiative of the Italian Ministry of University
(MUR) for the National Research Program (PNR). This project has been funded within the program
“PNRR Missione 4—Componente 2—Investimento 1.1 Fondo per il Programma Nazionale di Ricerca
e Progetti di Rilevante Interesse Nazionale (PRIN)”.}
\subsection{Authors Contributions}
Conceptualization, R.G., D.F. and N.T.Z.; Data curation, R.G.; Formal analysis, R.G., H.J., D.F. and N.T.Z.; Funding acquisition, D.F. and N.T.Z.; Investigation, R.G.; Methodology, R.G., H.J., D.F. and N.T.Z.; Project administration, H.J., D.F. and N.T.Z.; Software, R.G.; Supervision, H.J., D.F. and N.T.Z.; Validation, H.J., D.F. and N.T.Z.; Visualization, R.G; Writing – original draft, R.G., H.J., D.F. and N.T.Z.; Writing – review $\&$ editing, R.G., H.J., D.F. and N.T.Z.;
\subsection{Declaration of interests}
The authors declare no competing interests.

\bibliographystyle{unsrt}
\bibliography{sn-bibliography}

\begin{thebibliography}{99}
\bibitem[1]{De16}
		De~las Cuevas, G.; Cubitt, T.S.
		\newblock Simple universal models capture all classical spin physics.
		\newblock \href{https://doi.org/10.1126/science.aab3326}{\newblock{\em Science} {\bf 2016}, {\em 351},~1180--1183}.
		
		\bibitem[2]{Auerbach12}
		Auerbach, A.
		\newblock {\em Interacting Electrons and Quantum Magnetism}; Springer Science 
		\& Business Media: Berlin/Heidelberg, Germany,
		2012.
		
		\bibitem[3]{Wysin15}
		Wysin, G.M.
		\newblock Magnetism theory: spin models. In {\em Magnetic Excitations and 
		Geometric Confinement}; 2053--2563
		;
		\newblock \href{https://doi.org/10.1088/978-0-7503-1074-1ch2}{IOP Publishing: Bristol, UK, 2015; pp. 2--1 to 2--47}.
		
		\bibitem[4]{cialone2017tailoring}
		Cialone, M.; Celegato, F.; Co{\"\i}sson, M.; Barrera, G.; Fiore, G.; Shvab, R.; Klement, 
		U.; Rizzi, P.; Tiberto, P.
		\newblock Tailoring magnetic properties of multicomponent layered structure via 
		current annealing in FePd thin films.
		\newblock \href{https://doi.org/10.1038/s41598-017-16963-5}{{\em Sci. Rep.} {\bf 2017}, {\em 7},~16691}.
		
		\bibitem[5]{cialone2020comparative}
		Cialone, M.; Fernandez-Barcia, M.; Celegato, F.; Coisson, M.; Barrera, G.; Uhlemann, 
		M.; Gebert, A.; Sort, J.; Pellicer, E.; Rizzi, P.;  et~al.
		\newblock A comparative study of the influence of the deposition technique 
		(electrodeposition versus sputtering) on the properties of nanostructured Fe70Pd30 
		films.
		\newblock \href{https://doi.org/10.1080/14686996.2020.1780097}{{\em Sci. Technol. Adv. Mater.} {\bf 2020}, {\em 21},~424--434}.
		
		\bibitem[6]{Manousakis91}
		Manousakis, E.
		\newblock The spin-\textonehalf{} Heisenberg antiferromagnet on a square lattice 
		and its application to the cuprous oxides.
		\newblock \href{https://doi.org/10.1103/RevModPhys.63.1}{{\em Rev. Mod. Phys.} {\bf 1991}, {\em 63},~1--62}.
		
		\bibitem[7]{Bramwell98}
		Bramwell, S.; Harris, M.
		\newblock Frustration in Ising-type spin models on the pyrochlore lattice.
		\newblock \href{https://doi.org/10.1088/0953-8984/10/14/002}{{\em J. Physics: Condens. Matter.} {\bf 1998}, {\em 10},~L215}.
		
		\bibitem[8]{Mydosh_2015}
		Mydosh, J.A.
		\newblock Spin glasses: redux: an updated experimental/materials survey.
		\newblock \href{https://doi.org/10.1088/0034-4885/78/5/052501}{{\em Rep. Prog. Phys.} {\bf 2015}, {\em 78},~052501}.
		
		\bibitem[9]{Pan17}
		Pan, F.; Chico, J.; Delin, A.; Bergman, A.; Bergqvist, L.
		\newblock Extended spin model in atomistic simulations of alloys.
		\newblock \href{https://doi.org/10.1103/PhysRevB.95.184432}{{\em Phys. Rev. B} {\bf 2017}, {\em 95},~184432}.
		
		\bibitem[10]{Cavaliere2023}
		Cavaliere, F.; Razzoli, L.; Carrega, M.; Benenti, G.; Sassetti, M.
		\newblock Hybrid quantum thermal machines with dynamical couplings.
		\newblock \href{https://doi.org/10.1016/j.isci.2023.106235}{{\em iScience} {\bf 2023}, {\em 26}, 106235}.
		
		\bibitem[11]{Eckhardt2022}
		Eckhardt, C.J.; Passetti, G.; Othman, M.; Karrasch, C.; Cavaliere, F.; Sentef, M.A.; 
		Kennes, D.M.
		\newblock Quantum Floquet engineering with an exactly solvable tight-binding chain 
		in a cavity.
		\newblock \href{https://doi.org/10.1038/s42005-022-00880-9}{{\em Commun. Phys.} {\bf 2022}, {\em 5},~122}.
		
		\bibitem[12]{XiaoYong07}
		Feng, X.Y.; Zhang, G.M.; Xiang, T.
		\newblock Topological Characterization of Quantum Phase Transitions in a 
		Spin-$1/2$ Model.
		\newblock \href{https://doi.org/10.1103/PhysRevLett.98.087204}{{\em Phys. Rev. Lett.} {\bf 2007}, {\em 98},~087204}.
		
		\bibitem[13]{Duivenvoorden13}
		{Duivenvoorden, K.; Quella, T.
			\newblock Topological phases of spin chains.
			\newblock \href{https://doi.org/10.1103/PhysRevB.87.125145}{{\em Phys. Rev. B} {\bf 2013}, {\em 87},~125145}.
		
		\bibitem[14]{Dubinkin19}
		Dubinkin, O.; Hughes, T.L.
		\newblock Higher-order bosonic topological phases in spin models.
		\newblock \href{https://doi.org/10.1103/PhysRevB.99.235132}{{\em Phys. Rev. B} {\bf 2019}, {\em 99},~235132}.
		
		\bibitem[15]{Balents10}
		Balents, L.
		\newblock Spin liquids in frustrated magnets.
		\newblock \href{https://doi.org/10.1038/nature08917}{{\em Nature} {\bf 2010}, {\em 464},~199--208}.
		
		\bibitem[16]{Diep13}
		Diep, H.T.
		\newblock {\em Frustrated Spin Systems}; World Scientific: Singapore, 2013.
		
		\bibitem[17]{Sacco24}
		Sacco~Shaikh, D.; Catalano, A.G.; Cavaliere, F.; Franchini, F.; Sassetti, M.; 
		Traverso~Ziani, N.
		\newblock Phase diagram of the topologically frustrated XY chain.
        \newblock
        \href{https://link.springer.com/article/10.1140/epjp/s13360-024-05534-z}{{\em  Eur. Phys. J. Plus} {\bf 2024}, {\em 139},~1--14}
		
		\bibitem[18]{maric2020quantum}
		Mari{\'c}, V.; Giampaolo, S.M.; Franchini, F.
		\newblock Quantum phase transition induced by topological frustration.
		\newblock \href{https://doi.org/10.1038/s42005-020-00486-z}{{\em Commun. Phys.} {\bf 2020}, {\em 3},~220}.
		
		\bibitem[19]{lacroix2011introduction}
		Lacroix, C.; Mendels, P.; Mila, F.
		\newblock {\em Introduction to Frustrated Magnetism: Materials, Experiments, 
		Theory};  Springer Science \& Business Media: Berlin/Heidelberg, Germany,
		2011; Volume 164,
		
		\bibitem[20]{Mitra18}
		Mitra, A.
		\newblock Quantum quench dynamics.
		\newblock \href{https://doi.org/10.1146/annurev-conmatphys-031016-025451}{{\em Annu. Rev. Condens. Matter Phys.} {\bf 2018}, {\em 9},~245--259}.
		
		\bibitem[21]{Essler16}
		Essler, F.H.; Fagotti, M.
		\newblock Quench dynamics and relaxation in isolated integrable quantum spin 
		chains.
		\newblock \href{https://doi.org/10.1088/1742-5468/2016/06/064002}{{\em J. Stat. Mech. Theory Exp.} {\bf 2016}, {\em 2016},~064002}.
		
		\bibitem[22]{Porta18}
		Porta, S.; Gambetta, F.M.; Traverso~Ziani, N.; Kennes, D.M.; Sassetti, M.; Cavaliere, F.
		\newblock Nonmonotonic response and light-cone freezing in fermionic systems 
		under quantum quenches from gapless to gapped or partially gapped states.
		\newblock \href{https://doi.org/10.1103/PhysRevB.97.035433}{{\em Phys. Rev. B} {\bf 2018}, {\em 97},~035433}.
		
		\bibitem[23]{Porta20}
		Porta, S.; Cavaliere, F.; Sassetti, M.; Traverso~Ziani, N.
		\newblock Topological classification of dynamical quantum phase transitions in the 
		xy chain.
		\newblock \href{https://www.nature.com/articles/s41598-020-69621-8}{{\em Sci. Rep.} {\bf 2020}, {\em 10},~12766}.
		
		\bibitem[24]{Faure19}
		Faure, Q.; Takayoshi, S.; Simonet, V.; Grenier, B.; M\aa{}nsson, M.; White, J.S.; Tucker, 
		G.S.; R\"uegg, C.; Lejay, P.; Giamarchi, T.;  et~al.
		\newblock Tomonaga-Luttinger Liquid Spin Dynamics in the 
		Quasi-One-Dimensional Ising-Like Antiferromagnet 
		${\mathrm{BaCo}}_{2}{\mathrm{V}}_{2}{\mathrm{O}}_{8}$.
		\newblock \href{https://doi.org/10.1103/PhysRevLett.123.027204}{{\em Phys. Rev. Lett.} {\bf 2019}, {\em 123},~027204}.
		
		\bibitem[25]{faure2018topological}
		Faure, Q.; Takayoshi, S.; Petit, S.; Simonet, V.; Raymond, S.; Regnault, L.P.; Boehm, M.; 
		White, J.S.; M{\aa}nsson, M.; R{\"u}egg, C.;  et~al.
		\newblock Topological quantum phase transition in the Ising-like antiferromagnetic 
		spin chain BaCo2V2O8.
		\newblock \href{https://doi.org/10.1038/s41567-018-0126-8}{{\em Nat. Phys.} {\bf 2018}, {\em 14},~716--722}.
		
		\bibitem[26]{kinoshita2006quantum}
		Kinoshita, T.; Wenger, T.; Weiss, D.S.
		\newblock A quantum Newton's cradle.
		\newblock
        \href{https://doi.org/10.1038/nature04693}{{\em Nature} {\bf 2006}, {\em 440},~900--903}.
		
		\bibitem[27]{Holstein40}
		Holstein, T.; Primakoff, H.
		\newblock Field Dependence of the Intrinsic Domain Magnetization of a Ferromagnet.
		\newblock \href{https://doi.org/10.1103/PhysRev.58.1098}{{\em Phys. Rev.} {\bf 1940}, {\em 58},~1098--1113}.
		
		\bibitem[28]{popov1988functional}
		Popov, V.N.; Fedotov, S.
		\newblock The functional-integration method and diagram technique for spin 
		systems.
		\newblock \href{https://www.semanticscholar.org/paper/The-functional-integration-method-and-diagram-for/ca4a2fea661c6f7b037a8549c85b0dc722d8e4ff}{{\em Zh. Eksp. Teor. Fiz} {\bf 1988}, {\em 94},~183--194}.
		
		\bibitem[29]{Traverso23clock}
		Traverso, S.; Fleckenstein, C.; Sassetti, M.; Ziani, N.T.
		\newblock {An exact local mapping from clock-spins to fermions}.
		\newblock \href{https://doi.org/10.21468/SciPostPhysCore.6.3.055}{{\em SciPost Phys. Core} {\bf 2023}, {\em 6},~055}.
		
		\bibitem[30]{Ziani17}
		Traverso~Ziani, N.; Fleckenstein, C.; Dolcetto, G.; Trauzettel, B.
		\newblock Fractional charge oscillations in quantum dots with quantum spin Hall 
		effect.
		\newblock \href{https://doi.org/10.1103/PhysRevB.95.205418}{{\em Phys. Rev. B} {\bf 2017}, {\em 95},~205418}.
		
		\bibitem[31]{rodriguez2020relaxation}
		Rodriguez, R.; Parmentier, F.; Ferraro, D.; Roulleau, P.; Gennser, U.; Cavanna, A.; 
		Sassetti, M.; Portier, F.; Mailly, D.; Roche, P.
		\newblock Relaxation and revival of quasiparticles injected in an interacting quantum 
		Hall liquid.
		\newblock \href{https://doi.org/10.1038/s41467-020-16331-4}{{\em Nat. Commun.} {\bf 2020}, {\em 11},~2426}.
		
		\bibitem[32]{Gambetta15}
		Gambetta, F.M.; Ziani, N.T.; Barbarino, S.; Cavaliere, F.; Sassetti, M.
		\newblock Anomalous Friedel oscillations in a quasihelical quantum dot.
		\newblock \href{https://doi.org/10.1103/PhysRevB.91.235421}{{\em Phys. Rev. B} {\bf 2015}, {\em 91},~235421}.
		
		\bibitem[33]{TraversoZiani_2013}
		Ziani, N.T.; Cavaliere, F.; Sassetti, M.
		\newblock Theory of the STM detection of Wigner molecules in spin-incoherent 
		CNTs.
		\newblock \href{https://doi.org/10.1209/0295-5075/102/47006}{{\em Europhysics Letters} {\bf 2013}, {\em 102},~47006}.
		
		\bibitem[34]{cryst11010020}
		Ziani, N.T.; Cavaliere, F.; Becerra, K.G.; Sassetti, M.
		\newblock A Short Review of One-Dimensional Wigner Crystallization.
		\newblock \href{https://doi.org/10.3390/cryst11010020}{{\em Crystals} {\bf 2021}, {\em 11}}.
		
		\bibitem[35]{Bloch08}
		Bloch, I.; Dalibard, J.; Zwerger, W.
		\newblock Many-body physics with ultracold gases.
		\newblock \href{https://doi.org/10.1103/RevModPhys.80.885}{{\em Rev. Mod. Phys.} {\bf 2008}, {\em 80},~885--964}.
		
		\bibitem[36]{Cazalilla11}
		Cazalilla, M.A.; Citro, R.; Giamarchi, T.; Orignac, E.; Rigol, M.
		\newblock One dimensional bosons: From condensed matter systems to ultracold 
		gases.
		\newblock \href{https://doi.org/10.1103/RevModPhys.83.1405}{{\em Rev. Mod. Phys.} {\bf 2011}, {\em 83},~1405--1466}.
		
		\bibitem[37]{trotzky2012probing}
		Trotzky, S.; Chen, Y.A.; Flesch, A.; McCulloch, I.P.; Schollw{\"o}ck, U.; Eisert, J.; Bloch, 
		I.
		\newblock Probing the relaxation towards equilibrium in an isolated strongly 
		correlated one-dimensional Bose gas.
		\newblock \href{https://doi.org/10.1038/nphys2232}{{\em Nat. Phys.} {\bf 2012}, {\em 8},~325--330}.
		
		\bibitem[38]{Imambekov09}
		Imambekov, A.; Glazman, L.I.
		\newblock Universal Theory of Nonlinear Luttinger Liquids.
		\newblock \href{https://doi.org/10.1126/science.1165403}{{\em Science} {\bf 2009}, {\em 323},~228--231}.
		
		\bibitem[39]{Imambekov12}
		Imambekov, A.; Schmidt, T.L.; Glazman, L.I.
		\newblock One-dimensional quantum liquids: Beyond the Luttinger liquid paradigm.
		\newblock \href{https://doi.org/10.1103/RevModPhys.84.1253}{{\em Rev. Mod. Phys.} {\bf 2012}, {\em 84},~1253--1306}.
		
		\bibitem[40]{Wu06}
		Wu, C.; Bernevig, B.A.; Zhang, S.C.
		\newblock Helical Liquid and the Edge of Quantum Spin Hall Systems.
		\newblock \href{https://doi.org/10.1103/PhysRevLett.96.106401}{{\em Phys. Rev. Lett.} {\bf 2006}, {\em 96},~106401}.
		
		\bibitem[41]{Fiete06}
		Fiete, G.A.; Le~Hur, K.; Balents, L.
		\newblock Coulomb drag between two spin-incoherent Luttinger liquids.
		\newblock \href{https://doi.org/10.1103/PhysRevB.73.165104}{{\em Phys. Rev. B} {\bf 2006}, {\em 73},~165104}.
		
		\bibitem[42]{Fiete07}
		Fiete, G.A.
		\newblock Colloquium: The spin-incoherent Luttinger liquid.
		\newblock \href{https://doi.org/10.1103/RevModPhys.79.801}{{\em Rev. Mod. Phys.} {\bf 2007}, {\em 79},~801--820}.
		
		\bibitem[43]{Matveev07}
		Matveev, K.A.; Furusaki, A.; Glazman, L.I.
		\newblock Bosonization of strongly interacting one-dimensional electrons.
		\newblock \href{https://doi.org/10.1103/PhysRevB.76.155440}{{\em Phys. Rev. B} {\bf 2007}, {\em 76},~155440}.
		
		\bibitem[44]{JVoit_1995}
		Voit, J.
		\newblock One-dimensional Fermi liquids.
		\newblock \href{https://doi.org/10.1088/0034-4885/58/9/002}{{\em Rep. Prog. Phys.} {\bf 1995}, {\em 58},~977}.
		
		\bibitem[45]{Haldane81}
		Haldane, F.D.M.
		\newblock Effective Harmonic-Fluid Approach to Low-Energy Properties of 
		One-Dimensional Quantum Fluids.
		\newblock \href{https://doi.org/10.1103/PhysRevLett.47.1840}{{\em Phys. Rev. Lett.} {\bf 1981}, {\em 47},~1840--1843}.
		
		\bibitem[46]{Haldane_1981}
		Haldane, F.D.M.
		\newblock 'Luttinger liquid theory' of one-dimensional quantum fluids. I. Properties 
		of the Luttinger model and their extension to the general 1D interacting spinless Fermi 
		gas.
		\newblock \href{https://doi.org/10.1088/0022-3719/14/19/010}{{\em J. Phys. Solid State Phys.} {\bf 1981}, {\em 14},~2585}.
		
		\bibitem[47]{giamarchibook}
		Giamarchi, T.
		\newblock {\em Quantum Physics in One Dimension}; Oxford University Press: 
		Oxford, UK,  2003.
		
		\bibitem[48]{bethe1931theorie}
		Bethe, H.
		\newblock Zur theorie der metalle: I. Eigenwerte und eigenfunktionen der linearen 
		atomkette.
		\newblock \href{https://doi.org/10.1007/BF01341708}{{\em Zeitschrift f{\"u}r Physik} {\bf 1931}, {\em 71},~205--226}.
		
		\bibitem[49]{Jordan28}
		Jordan, P.; Wigner, E.
		\newblock {\"U}ber das Paulische {\"A}quivalenzverbot.
		\newblock \href{https://doi.org/10.1007/BF01331938}{{\em Z. Phys.} {\bf 1928}, {\em 47},~631--651}.
		
		\bibitem[50]{Franchini17}
		Franchini, F.
		\newblock {\em An Introduction to Integrable Techniques for One-Dimensional 
		Quantum Systems};  Springer: Berlin/Heidelberg, Germany,
		2017; Volume 940.
		
		\bibitem[51]{Bayat22}
		Bayat, A.; Bose, S.; Johannesson, H.
		\newblock {\em Entanglement in Spin Chains: From Theory to Quantum Technology 
		Applications}; Springer: Berlin/Heidelberg, Germany,
		2022.
		
		\bibitem[52]{Le_2018}
		Le, T.P.; Levinsen, J.; Modi, K.; Parish, M.M.; Pollock, F.A.
		\newblock Spin-chain model of a many-body quantum battery.
		\newblock \href{https://doi.org/10.1103/PhysRevA.97.022106}{{\em Phys. Rev. A} {\bf 2018}, {\em 97},~022106}.
		
		\bibitem[53]{Liu21}
		Liu, J.X.; Shi, H.L.; Shi, Y.H.; Wang, X.H.; Yang, W.L.
		\newblock Entanglement and work extraction in the central-spin quantum battery.
		\newblock \href{https://doi.org/10.1103/PhysRevB.104.245418}{{\em Phys. Rev. B} {\bf 2021}, {\em 104},~245418}.
		
		\bibitem[54]{Zhao21}
		Zhao, F.; Dou, F.Q.; Zhao, Q.
		\newblock Quantum battery of interacting spins with environmental noise.
		\newblock \href{https://doi.org/10.1103/PhysRevA.103.033715}{{\em Phys. Rev. A} {\bf 2021}, {\em 103},~033715}.
		
		\bibitem[55]{Catalano23}
		Catalano, A.; Giampaolo, S.; Morsch, O.; Giovannetti, V.; Franchini, F.
		\newblock Frustrating Quantum Batteries.
		\newblock \href{https://doi.org/10.1103/PRXQuantum.5.030319}{{\em PRX Quantum} {\bf 2024}, {\em 5},~030319}.
		
		\bibitem[56]{Grazi24}
		Grazi, R.; Sacco~Shaikh, D.; Sassetti, M.; Traverso~Ziani, N.; Ferraro, D.
		\newblock Controlling Energy Storage Crossing Quantum Phase Transitions in an 
		Integrable Spin Quantum Battery.
		\newblock \href{https://doi.org/10.1103/PhysRevLett.133.197001}{{\em Phys. Rev. Lett.} {\bf 2024}, {\em 133},~197001}.
		
		\bibitem[57]{Ali24SuperExt}
		Ali, A.; Elghaayda, S.; Al-Kuwari, S.; Hussain, M.I.; Rahim, M.T.; Kuniyil, H.; Seuda, 
		C.; Allati, A.E.; Mansour, M.; Haddadi, S.
		\newblock Super-Extensive Scaling in 1D Spin$-1/2$ $XY-\Gamma(\gamma)$ Chain 
		Quantum Battery.
		\emph{arXiv}  \textbf{2024},  \href{https://doi.org/10.48550/arXiv.2411.14074}{arXiv:2411.14074}.
		
		\bibitem[58]{Alicki_2013}
		Alicki, R.; Fannes, M.
		\newblock Entanglement boost for extractable work from ensembles of quantum 
		batteries. \newblock
        \href{https://doi.org/10.1103/PhysRevE.87.042123}{{\em Phys. Rev. E} {\bf 2013}, {\em 87},~042123}.
		
		\bibitem[59]{Bhattacharjee21}
		Bhattacharjee, S.; Dutta, A.
		\newblock Quantum thermal machines and batteries.
		\newblock \href{https://doi.org/10.1140/epjb/s10051-021-00235-3}{{\em  Eur. Phys. J. } {\bf 2021}, {\em 94}}.
		
		\bibitem[60]{Campaioli23}
		Campaioli, F.; Gherardini, S.; Quach, J.Q.; Polini, M.; Andolina, G.M.
		\newblock Colloquium: Quantum batteries.
		\newblock \href{https://doi.org/10.1103/RevModPhys.96.031001}{{\em Rev. Mod. Phys.} {\bf 2024}, {\em 96},~031001}.
		
		\bibitem[61]{Quach23}
		Quach, J.; Cerullo, G.; Virgili, T.
		\newblock Quantum batteries: The future of energy storage?
		\newblock \href{https://doi.org/https://doi.org/10.1016/j.joule.2023.09.003}{{\em Joule} {\bf 2023}, {\em 7},~2195--2200}.
		
		\bibitem[62]{Benenti17}
		Benenti, G.; Casati, G.; Saito, K.; Whitney, R.S.
		\newblock Fundamental aspects of steady-state conversion of heat to work at the 
		nanoscale.
		\newblock Fundamental aspects of steady-state conversion of heat to work at the 
		nanoscale.
        \newblock \href{https://doi.org/https://doi.org/10.1016/j.physrep.2017.05.008}{{\em Phys. Rep.} {\bf 2017}, {\em 694},~1--124}.
		
		\bibitem[63]{Esposito09}
		Esposito, M.; Harbola, U.; Mukamel, S.
		\newblock Nonequilibrium fluctuations, fluctuation theorems, and counting statistics 
		in quantum systems.
		\newblock \href{https://doi.org/10.1103/RevModPhys.81.1665}{{\em Rev. Mod. Phys.} {\bf 2009}, {\em 81},~1665--1702}.
		
		\bibitem[64]{Campisi16}
		Campisi, M.; Fazio, R.
		\newblock Dissipation, correlation and lags in heat engines.
		\newblock \href{https://doi.org/10.1088/1751-8113/49/34/345002}{{\em J. Phys. Math. Theor.} {\bf 2016}, {\em 49},~345002}.
		
		\bibitem[65]{Vinjanampathy16}
		Vinjanampathy, S.; Anders, J.
		\newblock Quantum thermodynamics.
		\newblock 
		\href{https://doi.org/10.1080/00107514.2016.1201896}{{\em Contemp. Phys.} {\bf 2016}, {\em 57},~545--579}.
		
		\bibitem[66]{Potts24}
		Potts, P.P.
		\newblock Quantum Thermodynamics. \newblock \href{https://doi.org/10.48550/arXiv.2406.19206}{\emph{arXiv}  \textbf{2024},  
		arXiv:2406.19206}.
		
		
		\bibitem[67]{Hu_2022}
		Hu, C.K.; Qiu, J.; Souza, P.J.P.; Yuan, J.; Zhou, Y.; Zhang, L.; Chu, J.; Pan, X.; Hu, L.; Li, 
		J.;  et~al.
		\newblock Optimal charging of a superconducting quantum battery.
		\newblock \href{https://doi.org/10.1088/2058-9565/ac8444}{{\em Quantum Sci. Technol.} {\bf 2022}, {\em 7},~045018}.
		
		\bibitem[68]{Gemme23}
		Gemme, G.; Andolina, G.M.; Pellegrino, F.M.D.; Sassetti, M.; Ferraro, D.
		\newblock Off-Resonant Dicke Quantum Battery: Charging by Virtual Photons.
		\newblock \href{https://doi.org/10.3390/batteries9040197}{{\em Batteries} {\bf 2023}, {\em 9}, 197}.
		
		\bibitem[69]{Dou23}
		Dou, F.Q.; Yang, F.M.
		\newblock Superconducting transmon qubit-resonator quantum battery.
		\newblock \href{https://doi.org/10.1103/PhysRevA.107.023725}{{\em Phys. Rev. A} {\bf 2023}, {\em 107},~023725}.
		
		\bibitem[70]{Razzoli24}
		Razzoli, L.; Gemme, G.; Khomchenko, I.; Sassetti, M.; Ouerdane, H.; Ferraro, D.; 
		Benenti, G.
		\newblock Cyclic solid-state quantum battery: Thermodynamic characterization and 
		quantum hardware simulation.    \newblock \href{https://doi.org/10.1088/2058-9565/ad9ed4}{\emph{Quantum Sci. Technol.} \textbf{2025}, 10(1), 015064}.
		
		\bibitem[71]{Cavaliere24}
		Cavaliere, F.; Gemme, G.; Benenti, G.; Ferraro, D.; Sassetti, M.
		\newblock Dynamical blockade of a reservoir for optimal performances of a 
		quantum battery.  \href{https://doi.org/10.48550/arXiv.2407.16471}{\emph{arXiv} \textbf{2024},  arXiv:2407.16471}.
		
		\bibitem[72]{Chiribella21}
		Chiribella, G.; Yang, Y.; Renner, R.
		\newblock Fundamental Energy Requirement of Reversible Quantum Operations.
		\newblock \href{https://doi.org/10.1103/PhysRevX.11.021014}{{\em Phys. Rev. X} {\bf 2021}, {\em 11},~021014}.
		
		\bibitem[73]{Menta24}
		Menta, R.; Cioni, F.; Aiudi, R.; Polini, M.; Giovannetti, V.
		\newblock Globally driven superconducting quantum computing architecture.  
		\emph{arXiv} \textbf{2024},  arXiv:2407.01182.
		
		
		\bibitem[74]{Elyasi2024}
		Elyasi, S.N.; Rossi, M.A.C.; Genoni, M.G.
		\newblock Experimental simulation of daemonic work extraction in open quantum 
		batteries on a digital quantum computer  \emph{arXiv}, \textbf{2024}, 
		arXiv:2410.16567.
		
		\bibitem[75]{Binder_2015}
		Binder, F.C.; Vinjanampathy, S.; Modi, K.; Goold, J.
		\newblock Quantacell: powerful charging of quantum batteries.
		\newblock \href{https://doi.org/10.1088/1367-2630/17/7/075015}{{\em New J. Phys.} {\bf 2015}, {\em 17},~075015}.
		
		\bibitem[76]{Andolina18}
		Andolina, G.M.; Farina, D.; Mari, A.; Pellegrini, V.; Giovannetti, V.; Polini, M.
		\newblock Charger-mediated energy transfer in exactly solvable models for quantum 
		batteries.
		\newblock \href{https://doi.org/10.1103/PhysRevB.98.205423}{{\em Phys. Rev. B} {\bf 2018}, {\em 98},~205423}.
		
		\bibitem[77]{Crescente22}
		Crescente, A.; Ferraro, D.; Carrega, M.; Sassetti, M.
		\newblock Enhancing coherent energy transfer between quantum devices via a 
		mediator.
		\newblock \href{https://doi.org/10.1103/PhysRevResearch.4.033216}{{\em Phys. Rev. Res.} {\bf 2022}, {\em 4},~033216}.
		
		\bibitem[78]{PERK1975319}
		Perk, J.; Capel, H.; Zuilhof, M.; Siskens, T.
		\newblock On a soluble model of an antiferromagnetic chain with alternating 
		interactions and magnetic moments.
		\newblock \href{https://doi.org/https://doi.org/10.1016/0378-4371(75)90052-7}{{\em Phys. Stat. Mech. Its Appl.} {\bf 1975}, {\em 81},~319--348}.
		
		\bibitem[79]{wakatsuki2014fermion}
		Wakatsuki, R.; Ezawa, M.; Tanaka, Y.; Nagaosa, N.
		\newblock Fermion fractionalization to Majorana fermions in a dimerized Kitaev 
		superconductor.
		\newblock \href{https://doi.org/10.1103/PhysRevB.90.014505}{{\em Phys. Rev. B} {\bf 2014}, {\em 90},~014505}.
		
		\bibitem[80]{Ziani20}
		Ziani, N.T.; Fleckenstein, C.; Vigliotti, L.; Trauzettel, B.; Sassetti, M.
		\newblock From fractional solitons to Majorana fermions in a paradigmatic model of 
		topological superconductivity.
		\newblock \href{https://doi.org/10.1103/PhysRevB.101.195303}{{\em Phys. Rev. B} {\bf 2020}, {\em 101},~195303}.
		
		\bibitem[81]{Jackiw76}
		Jackiw, R.; Rebbi, C.
		\newblock Solitons with fermion number \textonehalf{}.
		\newblock \href{https://doi.org/10.1103/PhysRevD.13.3398}{{\em Phys. Rev. D} {\bf 1976}, {\em 13},~3398--3409}.
		
		\bibitem[82]{Kivelson82}
		Kivelson, S.; Schrieffer, J.R.
		\newblock Fractional charge, a sharp quantum observable.
		\newblock \href{https://doi.org/10.1103/PhysRevB.25.6447}{{\em Phys. Rev. B} {\bf 1982}, {\em 25},~6447--6451}.
		
		\bibitem[83]{Goldstone81}
		Goldstone, J.; Wilczek, F.
		\newblock Fractional Quantum Numbers on Solitons.
		\newblock \href{https://doi.org/10.1103/PhysRevLett.47.986}{{\em Phys. Rev. Lett.} {\bf 1981}, {\em 47},~986--989}.
		
		\bibitem[84]{Heeger88}
		Heeger, A.J.; Kivelson, S.; Schrieffer, J.R.; Su, W.P.
		\newblock Solitons in conducting polymers.
		\newblock \href{https://doi.org/10.1103/RevModPhys.60.781}{{\em Rev. Mod. Phys.} {\bf 1988}, {\em 60},~781--850}.
		
		\bibitem[85]{qi2008fractional}
		Qi, X.L.; Hughes, T.L.; Zhang, S.C.
		\newblock Fractional charge and quantized current in the quantum spin Hall state.
		\newblock \href{https://doi.org/10.1038/nphys913}{{\em Nat. Phys.} {\bf 2008}, {\em 4},~273--276}.
		
		\bibitem[86]{Fleckenstein21}
		Fleckenstein, C.; Ziani, N.T.; Calzona, A.; Sassetti, M.; Trauzettel, B.
		\newblock Formation and detection of Majorana modes in quantum spin Hall 
		trenches.
		\newblock \href{https://doi.org/10.1103/PhysRevB.103.125303}{{\em Phys. Rev. B} {\bf 2021}, {\em 103},~125303}.
		
		\bibitem[87]{traverso2024emerging}
		Traverso, S.; Sassetti, M.; Traverso~Ziani, N.
		\newblock Emerging topological bound states in Haldane model zigzag nanoribbons.
		\newblock \href{https://doi.org/10.1038/s41535-023-00615-1}{{\em npj Quantum Mater.} {\bf 2024}, {\em 9},~9}.
		
		\bibitem[88]{Traverso22}
		Traverso, S.; Traverso~Ziani, N.; Sassetti, M.
		\newblock Effects of the Vertices on the Topological Bound States in a 
		Quasicrystalline Topological Insulator.
		\newblock \href{https://doi.org/10.3390/sym14081736}{{\em Symmetry} {\bf 2022}, {\em 14}}.
		
		\bibitem[89]{Traverso22role}
		Traverso, S.; Sassetti, M.; Ziani, N.T.
		\newblock Role of the edges in a quasicrystalline Haldane model.
		\newblock \href{https://doi.org/10.1103/PhysRevB.106.125428}{{\em Phys. Rev. B} {\bf 2022}, {\em 106},~125428}.
		
		\bibitem[90]{AYuKitaev_2001}
		Kitaev, A.Y.
		\newblock Unpaired Majorana fermions in quantum wires.
		\newblock \href{https://doi.org/10.1070/1063-7869/44/10S/S29}{{\em Physics-Uspekhi} {\bf 2001}, {\em 44},~131}.
		
		\bibitem[91]{Deng_2016}
		Deng, M.T.; Vaitiekėnas, S.; Hansen, E.B.; Danon, J.; Leijnse, M.; Flensberg, K.; 
		Nygård, J.; Krogstrup, P.; Marcus, C.M.
		\newblock Majorana bound state in a coupled quantum-dot hybrid-nanowire system.
		\newblock 
        \href{https://10.1126/science.aaf3961}{{\em Science} {\bf 2016}, {\em 354},~1557--1562}.
		
		\bibitem[92]{Prada_2018}
		Pe\~naranda, F.; Aguado, R.; San-Jose, P.; Prada, E.
		\newblock Quantifying wave-function overlaps in inhomogeneous Majorana 
		nanowires.
		\newblock \href{https://doi.org/10.1103/PhysRevB.98.235406}{{\em Phys. Rev. B} {\bf 2018}, {\em 98},~235406}.
		
		\bibitem[93]{Fleckenstein_2018}
		Fleckenstein, C.; Dom\'{\i}nguez, F.; Traverso~Ziani, N.; Trauzettel, B.
		\newblock Decaying spectral oscillations in a Majorana wire with finite coherence 
		length.
		\newblock \href{https://doi.org/10.1103/PhysRevB.97.155425}{{\em Phys. Rev. B} {\bf 2018}, {\em 97},~155425}.
		
		\bibitem[94]{Dibyendu_2013}
		Roy, D.; Bondyopadhaya, N.; Tewari, S.
		\newblock Topologically trivial zero-bias conductance peak in semiconductor 
		Majorana wires from boundary effects.
		\newblock \href{https://doi.org/10.1103/PhysRevB.88.020502}{{\em Phys. Rev. B} {\bf 2013}, {\em 88},~020502}.
		
		\bibitem[95]{Flensberg_2010}
		Flensberg, K.
		\newblock Tunneling characteristics of a chain of Majorana bound states.
		\newblock \href{https://doi.org/10.1103/PhysRevB.82.180516}{{\em Phys. Rev. B} {\bf 2010}, {\em 82},~180516}.
		
		\bibitem[96]{Law_2009}
		Law, K.T.; Lee, P.A.; Ng, T.K.
		\newblock Majorana Fermion Induced Resonant Andreev Reflection.
		\newblock \href{https://doi.org/10.1103/PhysRevLett.103.237001}{{\em Phys. Rev. Lett.} {\bf 2009}, {\em 103},~237001}.
		
		\bibitem[97]{Prada_2020}
		Prada, E.; San-Jose, P.; de~Moor, M.W.A.; Geresdi, A.; Lee, E.J.H.; Klinovaja, J.; Loss, 
		D.; Nyg{\aa}rd, J.; Aguado, R.; Kouwenhoven, L.P.
		\newblock From Andreev to Majorana bound states in hybrid 
		superconductor--semiconductor nanowires.
		\newblock \href{https://doi.org/10.1038/s42254-020-0228-y}{{\em Nat. Rev. Phys.} {\bf 2020}, {\em 2},~575--594}.
		
		\bibitem[98]{Ivanov_2001}
		Ivanov, D.A.
		\newblock Non-Abelian Statistics of Half-Quantum Vortices in ${p}$-Wave 
		Superconductors.
		\newblock \href{https://doi.org/10.1103/PhysRevLett.86.268}{{\em Phys. Rev. Lett.} {\bf 2001}, {\em 86},~268--271}.
		
		\bibitem[99]{Marra_2022}
		Marra, P.
		\newblock Majorana nanowires for topological quantum computation.
		\newblock
        \href{https://pubs.aip.org/aip/jap/article/132/23/231101/2837934/Majorana-nanowires-for-topological-quantum}{{\em J. Appl. Phys.} {\bf 2022}, {\em 132},~231101}.
		
		\bibitem[100]{Leijnse_2012}
		Leijnse, M.; Flensberg, K.
		\newblock Introduction to topological superconductivity and Majorana fermions.
		\newblock \href{https://doi.org/10.1088/0268-1242/27/12/124003}{{\em Semicond. Sci. Technol.} {\bf 2012}, {\em 27},~124003}.
		
		\bibitem[101]{Nayak_2008}
		Nayak, C.; Simon, S.H.; Stern, A.; Freedman, M.; Das~Sarma, S.
		\newblock Non-Abelian anyons and topological quantum computation.
		\newblock \href{https://doi.org/10.1103/RevModPhys.80.1083}{{\em Rev. Mod. Phys.} {\bf 2008}, {\em 80},~1083--1159}.
		
		\bibitem[102]{Kitaev_2003}
		Kitaev, A.
		\newblock Fault-tolerant quantum computation by anyons.
		\newblock \href{https://doi.org/https://doi.org/10.1016/S0003-4916(02)00018-0}{{\em Ann. Phys.} {\bf 2003}, {\em 303},~2--30}.
		
		\bibitem[103]{Oreg_2010}
		Oreg, Y.; Refael, G.; von Oppen, F.
		\newblock Helical Liquids and Majorana Bound States in Quantum Wires.
		\newblock \href{https://doi.org/10.1103/PhysRevLett.105.177002}{{\em Phys. Rev. Lett.} {\bf 2010}, {\em 105},~177002}.
		
		\bibitem[104]{Lutchyn_2010}
		Lutchyn, R.M.; Sau, J.D.; Das~Sarma, S.
		\newblock Majorana Fermions and a Topological Phase Transition in 
		Semiconductor-Superconductor Heterostructures.
		\newblock \href{https://doi.org/10.1103/PhysRevLett.105.077001}{{\em Phys. Rev. Lett.} {\bf 2010}, {\em 105},~077001}.
		
		\bibitem[105]{Smacchia11}
		Smacchia, P.; Amico, L.; Facchi, P.; Fazio, R.; Florio, G.; Pascazio, S.; Vedral, V.
		\newblock Statistical mechanics of the cluster Ising model.
		\newblock \href{https://doi.org/10.1103/PhysRevA.84.022304}{{\em Phys. Rev. A} {\bf 2011}, {\em 84},~022304}.
		
		\bibitem[106]{Ding19}
		Ding, C.
		\newblock Phase transitions of a cluster Ising model.
		\newblock \href{https://doi.org/10.1103/PhysRevE.100.042131}{{\em Phys. Rev. E} {\bf 2019}, {\em 100},~042131}.
		
		\bibitem[107]{Susskind1977}
		Susskind, L.
		\newblock Lattice fermions.
		\newblock \href{https://doi.org/10.1103/PhysRevD.16.3031}{{\em Phys. Rev. D} {\bf 1977}, {\em 16},~3031--3039}.
		
		\bibitem[108]{SHAMIR199390}
		Shamir, Y.
		\newblock Chiral fermions from lattice boundaries.
		\newblock \href{https://doi.org/https://doi.org/10.1016/0550-3213(93)90162-I}{{\em Nucl. Phys. } {\bf 1993}, {\em 406},~90--106}.
		
		\bibitem[109]{Neyenhuis17}
		Neyenhuis, B.; Zhang, J.; Hess, P.W.; Smith, J.; Lee, A.C.; Richerme, P.; Gong, Z.X.; 
		Gorshkov, A.V.; Monroe, C.
		\newblock Observation of prethermalization in long-range interacting spin chains.
		\newblock
        \href{https://10.1126/sciadv.1700672}{{\em Sci. Adv.} {\bf 2017}, {\em 3},~e1700672}.
		
		\bibitem[110]{Knight89}
		Kim, M.S.; de~Oliveira, F.A.M.; Knight, P.L.
		\newblock Properties of squeezed number states and squeezed thermal states.
		\newblock \href{https://doi.org/10.1103/PhysRevA.40.2494}{{\em Phys. Rev. A} {\bf 1989}, {\em 40},~2494--2503}.
		
		\bibitem[111]{Puskarov16}
		Puskarov, T.; Schuricht, D.
		\newblock {Time evolution during and after finite-time quantum quenches in the 
		transverse-field Ising chain}.
		\newblock \href{https://doi.org/10.21468/SciPostPhys.1.1.003}{{\em SciPost Phys.} {\bf 2016}, {\em 1},~003}.
		
		\bibitem[112]{Cole13}
		Jeske, J.; Vogt, N.; Cole, J.H.
		\newblock Excitation and state transfer through spin chains in the presence of 
		spatially correlated noise.
		\newblock \href{https://doi.org/10.1103/PhysRevA.88.062333}{{\em Phys. Rev. A} {\bf 2013}, {\em 88},~062333}.
		
		\bibitem[113]{Takei18}
		Aftergood, J.; Takei, S.
		\newblock Noise in tunneling spin current across coupled quantum spin chains.
		\newblock \href{https://doi.org/10.1103/PhysRevB.97.014427}{{\em Phys. Rev. B} {\bf 2018}, {\em 97},~014427}.
		
		\bibitem[114]{Gangadharaiah21}
		Singh, M.; Gangadharaiah, S.
		\newblock Driven quantum spin chain in the presence of noise: Anti-Kibble-Zurek 
		behavior.
		\newblock \href{https://doi.org/10.1103/PhysRevB.104.064313}{{\em Phys. Rev. B} {\bf 2021}, {\em 104},~064313}.
		
		\bibitem[115]{Troyer05}
		Werner, P.; Troyer, M.; Sachdev, S.
		\newblock Quantum Spin Chains with Site Dissipation.
		\newblock
        \href{https://doi.org/10.1143/JPSJS.74S.67}{{\em J. Phys. Soc. Jpn.} {\bf 2005}, {\em 74},~67--70}.
		
		\bibitem[116]{HongGang15}
		Chen, C.; An, J.H.; Luo, H.G.; Sun, C.P.; Oh, C.H.
		\newblock Floquet control of quantum dissipation in spin chains.
		\newblock \href{https://doi.org/10.1103/PhysRevA.91.052122}{{\em Phys. Rev. A} {\bf 2015}, {\em 91},~052122}.
		
		\bibitem[117]{Giedke13}
		Schwager, H.; Cirac, J.I.; Giedke, G.
		\newblock Dissipative spin chains: Implementation with cold atoms and steady-state 
		properties.
		\newblock \href{https://doi.org/10.1103/PhysRevA.87.022110}{{\em Phys. Rev. A} {\bf 2013}, {\em 87},~022110}.
		
		\bibitem[118]{Katsura19}
		Shibata, N.; Katsura, H.
		\newblock Dissipative spin chain as a non-Hermitian Kitaev ladder.
		\newblock \href{https://doi.org/10.1103/PhysRevB.99.174303}{{\em Phys. Rev. B} {\bf 2019}, {\em 99},~174303}}.
\end{thebibliography}

\end{document}